\documentclass[prd,preprint]{revtex4}

\usepackage{graphicx}

\usepackage{epsfig}

\usepackage{amssymb}

\begin{document}


\title{Test of Universal Rise of Hadronic Total Cross Sections based on $\pi p$, $Kp$ and $\bar pp,pp$ Scatterings}

\author{Muneyuki Ishida}

\email[ishida@phys.meisei-u.ac.jp]{}




\affiliation{Department of Physics, School of Science and Engineering, Meisei University, Hino, Tokyo 191-8506, Japan}

\author{Keiji Igi}

\affiliation{Theoretical Physics Laboratory, RIKEN, Wako, Saitama 351-0198, Japan}

\date{\today}

\begin{abstract}

Recently there are several evidences of the hadronic total cross section 
$\sigma_{\rm tot}$ to be proportional to $B$~log$^2s$ consistent with the Froissart 
unitarity bound. The COMPETE collaborations have further assumed 
$\sigma_{\rm tot}\simeq B$~log$^2(s/s_0)+Z$ to extend its universal 
rise with the common value of $B$ and $s_0$ for all hadronic scatterings to 
reduce the number of adjustable parameters. 
The coefficient $B$ was suggested to be universal in the arguments 
of colour glass condensate (CGC) of QCD in recent years. 
There has been, however, no rigorous proof yet based only on QCD. 
We attempt to investigate the value of $B$ for $\pi^\mp p$, $K^\mp p$ and 
$\bar pp,pp$ scatterings respectively through the search for the simultaneous 
best fit to the experimental $\sigma_{\rm tot}$ and $\rho$ ratios at high energies. 
The $\sigma_{\rm tot}$ at the resonance and intermediate energy regions has also 
been exploited as a duality constraint based on the special form of 
finite-energy sum rule(FESR). 
We estimate the values of $B$, $s_0$ and $Z$ individually for $\pi^\mp p$, $K^\mp p$
and $\bar pp,pp$ scatterings without using the universality hypothesis. 
It turs out that the values of $B$ are mutually consistent within one standard deviation. 
It has to be stressed that we cannot obtain such a definite conclusion 
without the duality constraint. It is also interesting to note that 
the values of $Z$ for $\pi p$, $Kp$ and $\bar p(p)p$ approximately satisfy 
the ratio $2:2:3$ predicted by the quark model. 
The obtained value of $B$ for $\bar p(p)p$ is $B_{pp}=0.280\pm 0.015$mb, which predicts
$\sigma_{\rm tot}^{pp}=108.0\pm 1.9mb$ and $\rho^{pp}=0.131\pm 0.0025$ at the LHC energy $\sqrt s=14$TeV.  
\end{abstract}

\pacs{11.55.Hx \sep 13.85.Lg}



\maketitle


\section{Introduction}

Recently there are several evidences\cite{r1,r2,r3,r4,r5,r6} of 
the total cross section $\sigma_{\rm tot}$ in the $\pi p$ and $\bar p(p)p$
scatterings to be proportional to log$^2s$ in high energies, consistent with 
the Froissart unitarity bound\cite{r7,r8}. 
The COMPETE collaborations\cite{r2,r6} have further assumed 
$\sigma_{\rm tot}\simeq B$~log$^2(s/s_0)$ to extend its universal rise with 
a common value of $B$ for all the hadronic scatterings. 
The universality of the coefficient $B$ was expected in the papers\cite{r9}, 
and other theoretical supports\cite{r11,r12} based on the arguments describing 
deep inelastic scattering by gluon saturation in hadron light-cone wave function 
(the Colour Glass Condensate\cite{r13} of QCD) were given in recent years. 
There has been, however, no rigorous proof yet based on QCD.

Therefore, it is worthwhile to prove or disprove 
this universal rise of $\sigma_{\rm tot}$  even empirically. 
In the near future, the $pp$ total cross section $\sigma_{\rm tot}^{pp}$ 
in $\sqrt s=14$TeV will be measured at TOTEM\cite{rTOTEM} and the other 
experiment\cite{rALFA} in the LHC. 
Therefore, the value of $B$ for $\bar pp,pp$ scattering, $B_{pp}$, will be 
determined with good accuracy. 
On the other hand, the $\pi^- N$ total cross sections $\sigma_{\rm tot}^{\pi^- N}$  
have been measured only up to $k=610$GeV where $k$ is the laboratory momentum of $\pi$  
and it corresponds to $\sqrt s=33.8$GeV, by the SELEX collaboration\cite{r14}. 
Thus, one might doubt to obtain the value of $B$ for $\pi p$ scattering, $B_{\pi p}$, 
with reasonable accuracy. 

In the previous article\cite{r5}, we attacked this problem by 
comparing the value of $B_{pp}$ and $B_{\pi p}$ in a new light. 
We used the laboratory energy of the incident particle, denoted as $\nu$, instead of 
the center of mass energy squared, $s$. They are related through 
\begin{eqnarray}
s &=& 2M\nu + M^2 + m^2
\label{eq1}
\end{eqnarray}
with each other where $M$ is the mass of target proton and $m$
is the mass of incident particle: $m=\mu$(pion mass), $m=m_K$(kaon mass) and $m=M$  
for $\pi p$, $Kp$ and $\bar p(p)p$ scatterings, respectively.
The total cross section $\sigma_{\rm tot}$ is composed of 
crossing-even cross section $\sigma_{\rm tot}^{(+)}$ and 
crossing-odd cross section $\sigma_{\rm tot}^{(-)}$. 
Its definition will be given in Sec.2. 
The $\sigma_{\rm tot}^{(+)}$ is a sum of Reggeon component and non-Reggeon component and 
$\sigma_{\rm tot}^{(-)}$ is only made of Reggeon component corresponding to the 
vector meson trajectories. 
The Reggeon components become negligible in the high-energy region. 
Thus, the $\sigma_{\rm tot}$ in high energies is described only by 
the non-Reggeon component of $\sigma_{\rm tot}^{(+)}$, which is parametrized by
\begin{eqnarray}
\sigma_{\rm tot}^{(+)} & \simeq & \frac{4\pi}{m^2}
\left(  c_2{\rm log}^2\frac{\nu}{m} + c_1{\rm log}\frac{\nu}{m} + c_0 \right)\ \ \ .
\label{eq2}
\end{eqnarray} 
The coefficients $c_2,c_1,c_0$ are introduced in the respective
scatterings.

The equation (\ref{eq2}) with $c_2>0$ shows the shape of parabola 
as a function of log~$\nu$ with a minimum.
The $c_2$ parameter controls the rise of parabola in high-energy side.
We re-express Eq.~(\ref{eq2}) as
\begin{eqnarray}
\sigma_{\rm tot}^{(+)} &\simeq & Z^{ap} + B~{\rm  log}^2\frac{s}{s_0} 
\label{eq3}
\end{eqnarray}
with $a = \pi^+ , K^+, p$. By using the relation $s\simeq 2M\nu$ from 
Eq.~(\ref{eq1}) approximated in high energies,
the $c_2$ in Eq.~(\ref{eq2}) is related directly 
with the $B$ parameter in Eq.~(\ref{eq3}). Thus,  
we obtain the $B$ parameters  for the relevant processes individually. 
In the case of $\bar pp$, we have data for large values of log~$\nu$  
coming from SPS and Tevatron experiments. 
Thus we can determine the value of $c_2(pp)$ and thus $B_{pp}$ with good accuracy. 
On the other hand, in the case of $\pi^- p,\pi^+ p$ scatterings 
we have used rich information of the experimental 
$\sigma_{\rm tot}$ data in the low and intermediate energy regions through 
the finite-energy sum rule (FESR). 
We used the FESR as a constraint 
between high-energy parameters, and analyzed the $\pi^\mp p$  total cross sections
$\sigma_{\rm tot}^{\pi^\mp p}$  and the $\rho$ ratios $\rho^{\pi^\mp p}$,
the ratios of real to imaginary parts of the forward scattering amplitudes. 
Here we adopted the FESR with the integral region between 
$k =\overline{N_1}$ and $\overline{N_2}$\cite{r15,r16}.
The $k$ is the laboratory momentum of the incident particle 
which is related with $\nu$ by $k=\sqrt{\nu^2-m^2}$. 
The $k\simeq \nu$ in high-energy regions. 
This FESR required that the low-energy extension of the high-energy asymptotic formula 
should coincide, roughly speaking, with the average of experimental $\sigma_{\rm tot}$  
in the relevant region between $k=\overline{N_1}$ and $\overline{N_2}$. 
This requirement is called the FESR duality. 
We have already used \cite{r15,r16} this sum rule between $\overline{N_1}=10$GeV and 
$\overline{N_2}=20$GeV.  
The rich data in $k<10$GeV were not included in this case, however. 
The lower-energy data are included in the integral of $\sigma_{\rm tot}$, 
the more precisely determined is the sub-leading term, i.e., the $P^\prime$ term 
( the term with coefficient $\beta_{P^\prime}$ in Eq.~(\ref{eq6}) corresponding to 
$f_2(1275)$ trajectory), which is built in the sense of FESR\cite{r17,r18,r19} 
by the sum of direct channel resonances. 
Then, it helps to determine the non-leading term such as log~$\nu$ 
which then helps to determine the leading term like log$^2\nu$. 
Thus, for the $\pi p$ scatterings, we are able to extend maximally the energy regions 
of the input data to take $\overline{N_1}\le 10$GeV, so as to obtain the value of 
$B_{\pi p}$ as most accurately as possible.
   
It is to be noted that the $\bar pp$ scattering has open (meson) channels 
below $\bar pp$-threshold with $\nu < M$(corresponding to 
$\sqrt s < 2M$), and $\sigma_{\rm tot}^{(+)}$ diverges above 
the threshold($\nu > M$) because of the exothermic reactions. 
The $K^-p$ scattering has also open channels with $\nu < m_K$
(corresponding to $\sqrt s< M+m_K$).  
If we choose the value of $\overline{N_1}$ to be fairly larger than 
$m\ (m=M$ for $\bar pp$, $m=m_K$ for $K^-p$), 
we have no difficulty coming from open channels. 
Contrarily, there are no such effects in $\pi p$ scattering . 
Thus, by taking $\overline{N_1}$ as small as possible, 
we can take into account more resonances through FESR in order to obtain 
the low-energy extension from the high-energy side with good accuracy. 
To obtain a sufficiently small error of $B_{\pi p}$, it appears to be important to include 
the information of the low-energy scattering data with $0\le k \le 10$GeV through FESR. 
   
We will show that the resulting value of $B_{\pi p}$ is consistent\cite{r5} with 
that of $B_{pp}$, which appears to support the universality hypothesis. 
It will also be shown that the central value of $B_{Kp}$ is also consistent with $B_{pp}$  
and $B_{\pi p}$, although its error is fairly large, due to the present situation of 
the $K^-p,\ K^+p$  data. 
So far, we have searched for the simultaneous best fit of the high-energy parameters 
such as $c_2,c_1,c_0\cdots$ to the $\sigma_{\rm tot}$ and the $\rho$ ratios under 
the duality constraint. 
In other words, both $B$(related to $c_2$) and $s_0$(related to $c_1/c_2$) was 
completely arbitrary. 

We have also attempted to fit data by assuming the universality of $B$  in 
$\sigma_{\rm tot}\sim B$~log$^2(s/s_0)$ from the beginning.
The fit is successful and the increase of total fitting $\chi^2$ due to
the universality constraint is small. This result also suggests the universality of $B$.
The scale $s_0$ has been assumed to be independent of 
the colliding particles in ref.\cite{r6}. 
This resulted in reducing the number of adjustable parameters.
But there has been no proof again on this assumption based on QCD.
We will also investigate this possibility. 

In Sec.2, kinematical considerations are summarized for forward 
$\pi^\mp p$, $K^\mp p$ and $\bar p(p)p$ scatterings. 
A duality constraint is also explained based on the special form of FESR.
In Sec. 3, we explain the approach how to estimate the value of $B$, $s_0$ and $Z$
individually for the above hadron scatterings.
In Sec. 4, detailed analyses are given based on $\sigma_{\rm tot}$ and $\rho$ together
with the duality constraint. 
We then discuss about the universality of coefficient $B$.
Sec. 5 summarizes our conclusions.

\section{Kinematical considerations}

\subsection{Total cross sections $\sigma_{\rm tot}$ and $\rho$ ratios}

We take both the crossing-even and crossing-odd forward scattering amplitudes, 
$F^{(+)}(\nu )$ and $F^{(-)}(\nu )$, defined by
\begin{eqnarray}
F^{(\pm )}(\nu ) &=& \frac{f^{\bar ap}(\nu ) \pm f^{ap}(\nu )}{2}
\label{eq4}\\
f^{\bar ap}(\nu ) &=& F^{(+)}(\nu ) + F^{(-)}(\nu ) \nonumber\\
f^{ap}(\nu ) &=& F^{(+)}(\nu ) - F^{(-)}(\nu )
\nonumber
\end{eqnarray}
where $(\bar a,a)=(\pi^-,\pi^+)$, $(K^-,K^+)$ and $(\bar p,p)$ respectively, and 
$f^{\bar ap}(\nu );f^{ap}(\nu )$ is the forward $\bar ap;ap$ scattering amplitudes.
The $\nu$ is the incident energy of $\bar p(p),\ \pi$ and $K$ in the laboratory system. 
The combinations (\ref{eq4}) of the amplitudes satisfy the crossing property
\begin{eqnarray}
F^{(\pm )}(-\nu ) &=& \pm F^{(\pm )}(\nu )^*
\label{eq5}
\end{eqnarray}
under crossing transformation $\nu \rightarrow -\nu$ for forward amplitudes.
We assume that
\begin{eqnarray}
{\rm Im}F^{(+)}(\nu ) 
 & \simeq & \frac{\nu}{m^2}  \left( c_0  +  c_1 {\rm log}\frac{\nu }{m} 
     + c_2 {\rm log}^2\frac{\nu }{m}  \right)
    + \frac{\beta_{P^\prime}}{m} \left( \frac{\nu}{m}\right)^{\alpha_{P^\prime}}\ 
\label{eq6}\\
{\rm Im}F^{(-)}(\nu ) & \simeq &  
     \frac{\beta_V}{m} \left( \frac{\nu}{m}\right)^{\alpha_V}\ ,
\label{eq7}
\end{eqnarray}
for $\nu > N$ with some energy $N$ due to the Pomeron-Reggeon exchange model 
except for the terms with coefficients $c_2$  and $c_1$ . 
The coupling coefficients $\beta_{P^\prime},\ c_0,\ \beta_V$ are the unknown 
parameters in the Regge theory. 
The $\alpha_{P^\prime},\ \alpha_V$  are determined phenomenologically. 
The $c_2,\ c_1$ terms are introduced consistently with the Froissart bound 
to describe  the rise of $\sigma_{\rm tot}$ in the high-energy regions. 
The total cross sections $\sigma_{\rm tot}^{\bar ap},\ \sigma_{\rm tot}^{ap}$ 
and the $\rho$ ratios $\rho^{\bar ap}$ and $\rho^{ap}$ are given by 
\begin{eqnarray}
{\rm Im}\ f^{\bar ap,ap}(\nu ) &=& \frac{k}{4\pi}\sigma_{\rm tot}^{\bar ap,ap}\ , \nonumber\\
\rho^{\bar ap} &=& \frac{{\rm Re}\ f^{\bar ap}}{{\rm Im}\ f^{\bar ap}},\ \ \ 
\rho^{ap} = \frac{{\rm Re}\ f^{ap}}{{\rm Im}\ f^{ap}}, 
\label{eq8}
\end{eqnarray}
respectively, where the $k$ are the incident momenta of $\bar p(p),\ \pi$ 
and $K$ in the laboratory system. 
The total cross section of crossing-even(odd) part $\sigma_{\rm tot}^{(\pm )}$
is given by $\sigma_{\rm tot}^{(\pm )}=(\sigma_{\rm tot}^{\bar ap}\pm \sigma_{\rm tot}^{ap})/2
=\frac{4\pi}{k}{\rm Im}F^{(\pm )}(\nu )$.
By using the crossing property (\ref{eq5}), 
the real parts are given by\cite{r4,r16}
\begin{eqnarray}
{\rm Re}F^{(+)}(\nu ) & \simeq & \frac{\pi \nu}{2m^2}\left( 
c_1 + 2 c_2 {\rm ln}\frac{\nu}{m} \right)
  -\frac{\beta_{P^\prime}}{m}\left(\frac{\nu}{m}\right)^{\alpha_{P^\prime}}
  {\rm cot}\frac{\pi\alpha_{P^\prime}}{2} + F^{(+)}(0)\ , \label{eq9}\\
{\rm Re}F^{(-)}(\nu ) & \simeq & \frac{\beta_V}{m}\left(\frac{\nu}{m}\right)^{\alpha_V}
  {\rm tan}\frac{\pi\alpha_V}{2} \ ,
\label{eq10}
\end{eqnarray}
where $F^{(+)}(0)$ is a subtraction constant.

\subsection{Duality constraints}

The FESR is used as a duality constraint between these parameters\cite{r15,r16}.
\begin{eqnarray}
\frac{2}{\pi}\int_{N_1}^{N_2} && \frac{\nu }{k^2} 
 \ {\rm Im}F^{(+)}(\nu )\ d\nu 
  = \frac{1}{2\pi^2} \int_{\overline{N}_1}^{\overline{N}_2} 
\ \sigma^{(+)}_{\rm tot}(k )\ dk \ \ .\ \ \ \ \ \ \ 
\label{eq11}
\end{eqnarray}
The laboratory energy $\nu$ is related with corresponding momentum $k$ 
by $\nu =\sqrt{k^2+m^2}$. The momentum 
corresponding to $\nu =N$ is represented by the quantity with overline 
such as $k=\overline{N}$ in this paper. 
The $\nu =N_{1,2}$ in Eq.~(\ref{eq11}) are related to 
the corresponding momenta 
$k=\overline{N_{1,2}}$ by $N_{1,2}=\sqrt{\overline{N_{1,2}}^2+m^2}$.  
The value of $\overline{N_2}$ should be selected to be reasonably high momentum above 
which no resonance structures are observed, while $\overline{N_1}$ may be taken to be 
in the resonance region in the sense of the FESR duality.

The integrand of the LHS of Eq.~(\ref{eq11}) is the low-energy extension of 
Eq.~(\ref{eq6}). 
The RHS is  the integral of experimental 
$\sigma_{\rm tot}^{(+)}(=(\sigma_{\rm tot}^{\bar ap}+\sigma_{\rm tot}^{ap})/2)$  
in the resonance-energy regions. 
This shows up several peak and dip structures corresponding to a number of resonances, 
in addition to the non-resonating background. 
Thus, Eq.~(\ref{eq11}) means the FESR duality, that is, the average of these resonance
structures plus the non-resonating background in $\sigma_{\rm tot}^{(+)}$ should coincide with 
the low-energy extension of the asymptotic formula.

\section{The General Approach}

\subsection{Energy region of $\sigma_{\rm tot}$ and $\rho$ fitted by asymptotic formulas}

Let us first discuss the energy regions where experimental $\sigma_{\rm tot}$
and $\rho$ ratios can be fitted by the asymptotic forms (\ref{eq6})-(\ref{eq10}).

As is well known, many low-energy resonances smoothly join into the smooth high-energy 
behviours around the transition energy $\nu_0$. 
This energy $\nu_0$ are around 5GeV (which corresponds to $\sqrt s\simeq 3.3$GeV in
$\bar pp$ scattering) in real experimental data. 
This value of $\nu_0$ are in the energy region of overlapping resonances and
seems to be too small to apply the asymptotic formula to the data
just above $\nu = \nu_0$.
However, since the average of low-energy resonances is equivalent to 
the asymptotic formula due to the FESR duality,
we can equate the experimental $\sigma_{\rm tot}$ to the 
imaginary part of $F^{(\pm )}(\nu )$ (\ref{eq6}),(\ref{eq7}) for $\nu > \nu_0$.
Let us now consider the behaviours of the real part of the amplitude. 
As a simplicity of explanation, let us consider the crossing-odd amplitude. 
Then, we can substitute the RHS of (\ref{eq7}) into the principal part 
dispersion integral, instead of substituting low-energy resonances, 
due to the FESR duality. 
Therefore, we can obtain the RHS of (\ref{eq10}) explicitly for $\nu > \nu_0$.

This can be easily extended to the general case. 
When the scattering amplitude includes the contribution from the Pomeron exchange, 
the two-component hypothesis of duality was proposed by Gilman,Harari and Zarmi\cite{r20}, 
i.e., the ordinary Regge pole ( $P^\prime$ ) are built 
by direct-channel resonances in the sense of FESR, while the Pomeron-type 
singularity (which corresponds to Eq.~(\ref{eq2}) in the present case) is associated 
with the non-resonating background. 
If we take this hypothesis, the same argument can be applied, and we can substitute 
the RHS of (\ref{eq6}) into the principal-part dispersion relation from threshold to obtain 
the RHS of (\ref{eq9}). 
Therefore, we can use the RHS of (\ref{eq9}) for $\nu > \nu_0$. 
Since the transition energy $\nu_0$ is around 5GeV as mentioned above, 
we can use the asymptotic form for 
both the imaginary part and real part for  $k > 5$GeV.

\subsection{Practical approach for the search of $B$}

In order to obtain the value of $B$, we search for the simultaneous 
best fit to $\sigma_{\rm tot}^{(+)}$  and the $\rho^{(+)}$ ratios 
under the duality constraint, Eq.(\ref{eq11}).  
The formula (\ref{eq4})|(\ref{eq10}) and the duality constraint (\ref{eq11}) are our starting points. 
The LHS of Eq.~(\ref{eq11}) is the integral of the asymptotic formulas (\ref{eq6}) 
and is represented by a linear homogeneous equation of the parameters $c_{2,1,0}$ and $\beta_{P^\prime}$. 
The RHS of Eq.~(\ref{eq11}) is the integral of experimental $\sigma_{\rm tot}$, which is estimated 
by using the experimental data of $\sigma_{\rm tot}^{\pi^\mp p},\sigma_{\rm tot}^{K^\mp p},\sigma_{\rm tot}^{\bar pp,pp}$. 
Here $\overline{N_2}$ is fixed to be 20GeV while we take various values of $\overline{N_1}$ being less 
than 10GeV. We try to take $\overline{N_1}$ as small as possible.
The $\sigma_{\rm tot}^{\bar ap,ap}$  and  $\rho^{\bar ap,ap}$ are fitted simultaneously for the respective 
processes of $\pi p$, $Kp$ and $\bar p(p)p$  scatterings. 
In actual analyses we fit the data of Re $f^{\bar ap,ap}(k )$ by the formulas (\ref{eq9}) and (\ref{eq10}). 
We have made them from experimental data of 
$\rho^{\bar ap,ap}$ in ref.\cite{r6} times Im~$f^{\bar ap,ap}_{\rm PDG}(k )$
where we use the result of the fit given in PDG\cite{r6}\footnote{
In p.337 of ref.\cite{r6} $\sigma_{\rm tot}^{\bar ap,ap}$ is given by 
$\sigma_{\rm tot}^{\bar ap,ap}=Z^{ap}+B~{\rm log}^2\frac{s}{s_0}+Y_1^{ap}(\frac{s_1}{s})^{\eta_1}
\pm Y_2^{ap}(\frac{s_1}{s})^{\eta_2}$. The scale $s_1$ is fixed to be 1GeV$^2$. 
The $(B,\ \sqrt{s_0},\ \eta_1,\ \eta_2)=(0.308(10){\rm mb},\ 5.38(50){\rm GeV},\ 0.458(17),\ 0.545(7))$ 
obtained process-independently.
The process-dependent parameters $(Z^{ap},\ Y_1^{ap},\ Y_2^{ap})$ are obtained as 
(20.86(40), 19.24(1.22), 6.03(19)), (17.91(36), 7.1(1.5), 13.45(40)) and (35.45(48), 42.53(1.35), 33.34(1.04)) 
for $\pi^\mp p,\ K^\mp p$ and $\bar p(p)p$, respectively.
We made Im $f^{\bar ap,ap}_{\rm PDG}(\nu )$ by $\frac{k}{4\pi}\times \sigma_{\rm tot}^{\bar ap,ap}$ with the central values of 
these parameters.  Multiplying Im~$f^{\bar ap,ap}_{\rm PDG}(\nu )$ with 
the $\rho^{\bar ap,ap}$ data at the corresponding energies
we obatin the data of Re $f^{\bar ap,ap}$ . Here we omit errors of $\sigma_{\rm tot}^{\bar ap,ap}$ since
errors of $\rho^{\bar ap,ap} $ are generally larger than $\sigma_{\rm tot}^{\bar ap,ap}$.
}.
By using Re~$f^{\bar ap,ap}$ data in the fittings, the resulting $\chi^2$ functions
become second-order homogeneous equations of the relevant parameters, which are easy
handle. It makes our analyses simple and transparent.

The $\sigma_{\rm tot}^{\bar ap,ap}$  for $k\ge 20$GeV and  $\rho^{\bar ap,ap}$ for $k\ge 5$GeV are fitted 
simultaneously.
Here, by considering the transition energy $\nu_0\sim 5$GeV, we have chosen $k\ge 5$GeV for the fitted energy region 
of the $\rho$ data, while for $\sigma_{\rm tot}$, $k\ge 20$GeV, being different from the $\rho$ data.
The $\sigma_{\rm tot}$ data up to $k=20$GeV are used to obtain their integrals 
in the duality constraint (\ref{eq11}).
Thus, in order to avoid the double-counting of the data we use larger 
values $k > 20$GeV  for $\sigma$  than that for  $\rho$.

The high-energy parameters  $c_2,c_1,c_0,\beta_{P^\prime}$ and $\beta_V$ are treated as process-dependent, 
while $\alpha_{P^\prime}$  and $\alpha_V$ are fixed with common values for every process. 
The duality constraints (\ref{eq11}) give constraints between $c_2,c_1,c_0$  and $\beta_{P^\prime}$ for 
$\bar p(p)p,\ Kp$  and $\pi p$  scatterings, respectively.
The $F^{(+)}(0)$ is treated as an additional parameter, and the number of fitting parameters is five for each process. 
The resulting $c_2$ are related to the $B$ parameters, defined by $\sigma \simeq B {\rm log}^2(s/s_0)+\cdots$,
 through the equation
\begin{eqnarray}
B_{ap}  &=& \frac{4\pi}{m^2}c_2,\ \ \ m=M,\mu ,m_K,\ \ \ a=p,\pi ,K
\label{eq12}
\end{eqnarray}
and  we can test the universality of the $B$ parameters.

\subsection{Analysis when coefficient $B$,  scale $s_0$ are assumed to be universal}

We also analyze the data by assuming the coefficient $B$ of 
 $\sigma_{\rm tot}\sim B~{\rm log}^2(s/s_0)$ to be universal from the beginning
and test the universality of $B$.
We also search for the possibility that
the $s_0$ has a common value for $pp$, $\pi p$, $Kp$ scatterings.


\section{Analysis based on $\sigma_{\rm tot}$ and $\rho$}

\subsection{Evaluation of the integral of $\sigma_{\rm tot}^{(+)}$ appearing in FESR}

In order to obtain the explicit forms of FESR (\ref{eq11}),
it is necessary to evaluate the integral of $\sigma_{\rm tot}^{(+)}$,
\begin{eqnarray}
\frac{1}{2\pi^2}\int_{\overline{N_1}}^{\overline{N_2}} \sigma_{\rm tot}^{(+)}(k)\ dk
&=& \frac{1}{2\pi^2}\int_{\overline{N_1}}^{\overline{N_2}}
\frac{1}{2}( \sigma_{\rm tot}^{\bar ap}(k) + \sigma_{\rm tot}^{ap}(k) )\ dk\ \ \ \ \ 
\label{eq13}
\end{eqnarray}
with $a=\pi^+,K^+,p$ 
from the experimental data\cite{r6} for each process.
For this purpose we have performed phenomenological fits to the experimental
$\sigma_{\rm tot}^{\bar ap,ap}$.
The experimental $\sigma_{\rm tot}$ 
can be fitted simply by the phenomenological formula
$\sigma_{\rm tot}=\frac{4\pi}{k}\{ \frac{\nu}{m^2}
(c_2{\rm log}^2\frac{\nu}{m}+c_1{\rm log}\frac{\nu}{m}+c_0)
+\frac{\beta}{m}(\frac{\nu}{m})^{0.5} +\frac{d}{m}(\frac{\nu}{m})^{-0.5} 
+\frac{f}{m}(\frac{\nu}{m})^{-1.5} +\frac{g}{m}(\frac{\nu}{m})^{-2.5}\}$.
The dimensionless parameters $c_2$ are fixed to the values of our previous 
analysis\cite{r15}, $c_2=0.00140,\ 0.0185$ and $0.0520$ for 
$\pi^\mp p,\ K^\mp p$ and $\bar p(p)p$ scatterings, respectively.
The other parameters are taken to be free, dependent on the processes.
The error of each data point, denoted as $\Delta y$, is given by combining  
statistical error $\Delta y_{\rm stat}$ and systematic error $\Delta y_{syst}$ 
as $\Delta y\equiv \sqrt{(\Delta y_{\rm stat})^2+(\Delta y_{\rm syst})^2}$.
The data of $\sigma_{\rm tot}^{\bar pp}$ 
are mutually inconsistent in the low-energy region
among the data of different experiments, 
and we cannot obtain a good fit.
We adopt a statistical method, Sieve algorithm\cite{r21}, and  
seven points are removed following this prescription\footnote{
The following data points 
are removed: $(k$(GeV),$\sigma_{\rm tot}^{\bar pp}$(mb))$=
(2.5,79.4\pm 1.0),(3.54,69.7\pm 0.5),(3.6,76.2\pm 1.8),(4.,71.\pm 1.),
(4.015,66.84\pm 0.32),(4.3,60.6\pm 0.8),(9.14,57.51\pm 0.73)$.
This procedure is explained in ref.\cite{r16}.
}.
The experimental $\sigma_{\rm tot}^{\bar ap,ap}(k)$ are 
fitted in the region of laboratory momenta, $k_a <k<100$GeV, where 
$k_a$ are taken to be 3GeV, 2GeV and 2.5GeV for $\pi p,Kp$ and $\bar p(p)p$ scatterings, 
respectively. The $k_a$ correspond to 
$\sqrt s=2.56$GeV, 2.35GeV and 2.60GeV for the relevant processes. 
There are no remarkable resonance structures observed above these energies, 
and successful fits are obtained by using commonly this simple formula.
The $g$ is fixed to be $g=0$ for the analyses of $\pi^\mp p$ and $\bar p(p)p$, while 
it is treated to be free for $K^\mp p$.
The number of fitting parameters are 5 respectively for ${\pi^- p}$,
${\pi^+ p}$, ${\bar pp}$ and $pp$, while 6 for $K^-p$ and for $K^+p$.
The resulting $\chi^2$ are $\chi^2/(N_D-N_P)=102.6/(165-5),69.7/(100-5)$ 
for $\pi^-p,\pi^+p$; $171.4/(149-6),75.0/(86-6)$ for $K^-p,K^+p$; 
$48.8/(70-5),112.2/(103-5)$ for $\bar pp,pp$\footnote{The $\chi^2$=48.8 
for sieved $\bar pp$ data includes the 
factor $R=1.140$. See ref.\cite{r21}.}.

In the FESR (\ref{eq11}) the $\overline{N_2}$ is fixed to be
$\overline{N_2}=20$GeV, while we take various values of $\overline{N_1}$.
The integrals of $\sigma_{\rm tot}^{(+)}(k)$ from the relevant $\overline{N_1}$
to the $\overline{N_2}$ are estimated by using these phenomenological fits.

In $\pi p$ scatterings, we try to take very small values of 
$\overline{N_1}$ in the resonance-energy region.
There are no open channels below threshold $\nu <\mu$ in this process, and 
the smaller value of $\overline{N_1}$ is taken, the more information in the low-energy
region is included through FESR, and the more accurate value of $c_2$ is obtained. 
Actually we take $\overline{N_1}$ less than 1GeV.
In this case we divide the region of the integral into two parts:
The integral of $\sigma_{\rm tot}^{(+)}(k)$ 
from the higher-energy region, $k_d<k<\overline{N_2}$, is estimated by using 
the phenomenological fits, while the integral from the lower-energy region, 
$\overline{N_1}<k<k_d$, are evaluated directly from experimental data.
That is, the data points are connected by straight lines and the areas of these 
polygonal line graphs are regarded as the corresponding integrals. 
The dividing momentum $k_d$ is taken to be 4GeV.

\begin{figure}
\resizebox{0.75\textwidth}{!}{

  \includegraphics{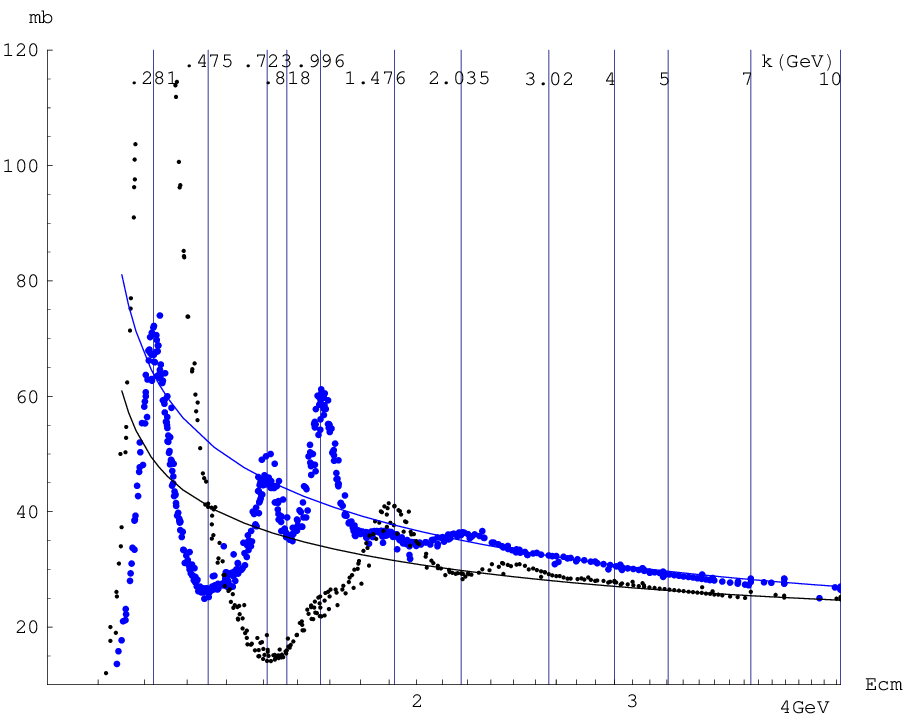}

}


\caption{The $\sigma_{\rm tot}$ data of $\pi^-p$ scattering(big blue points) and 
of $\pi^+p$ scattering(small black points) in the low-energy region:
The errors are not shown. 
The horizontal axis represents center of mass energy $E_{cm}$.
The vertical lines correspond to the values of 
laboratory momenta 
$k$=$\overline{N_1}$=10,7,5,4,3.02,2.035,1.476,0.9958,0.818,0.723,0.475,0.281 GeV,
which are selected as the lower limits of the integrals in FESR (\ref{eq11}).
Their numbers(in GeV) are shown in the upper side. 
The solid lines represent the low-energy extensions of our best fit using FESR 
in the case $\overline{N_1}=0.818$GeV.
}

\label{fig1}

\end{figure}

The $\sigma_{\rm tot}^{\pi^\mp p}$ data in low-energy region are shown 
in Fig.~\ref{fig1}.\\
We take
$k$=$\overline{N_1}$=10,7,5,4,3.02,2.035,1.476,0.9958,0.818,0.723,0.475,0.281 GeV.
These values of $\overline{N_1}$, which are shown by vertical lines, correspond to the 
laboratory momenta of peak and dip positions observed in experimental 
$\sigma_{\rm tot}^{\pi^\mp p}$ spectra. This can be seen in Fig.~\ref{fig1}. 

The values of cross-section integrals estimated from the above-mentioned procedures
are given in Table \ref{tab1}.

\begin{table}
\caption{The integral of cross section 
$\frac{1}{2\pi^2}\int_{\overline{N_1}}^{\overline{N_2}}
\sigma_{\rm tot}^{\pi^\mp p}(k)\ dk\ ({\rm GeV}^{-1})$
and their average estimated by using experimental data
for $\pi^\mp p$ scattering:
The $\overline{N_2}$ is fixed to be 20GeV while we take various values of  
$\overline{N_1}$.
}
\begin{tabular}{c|ccc}
$\overline{N_1}$--$\overline{N_2}$(GeV)
 & $\frac{1}{2\pi^2}\int_{\overline{N_1}}^{\overline{N_2}}
\sigma_{\rm tot}^{\pi^-p}(k)\ dk$
 & $\frac{1}{2\pi^2}\int_{\overline{N_1}}^{\overline{N_2}}
\sigma_{\rm tot}^{\pi^+p}(k)\ dk$
 & $\frac{1}{2\pi^2}\int_{\overline{N_1}}^{\overline{N_2}}
\sigma_{\rm tot}^{(+)}(k)\ dk$
  \\
\hline
10--20 & {31.404$\pm$0.033} & {33.611$\pm$0.029} & {32.508$\pm$0.022}\\
 7--20 & 41.253$\pm$0.042 & 44.244$\pm$0.038 & {42.748$\pm$0.028}\\
 5--20 & 48.069$\pm$0.047 & 51.656$\pm$0.044 & {49.863$\pm$0.032}\\
 4--20 & 51.609$\pm$0.048 & 55.536$\pm$0.047 & {53.572$\pm$0.034}\\
 3.02--20 & {55.220$\pm$0.050} & {59.539$\pm$0.052} & {57.380$\pm$0.036}\\
2.035-20 & {59.069$\pm$0.052} & {63.899$\pm$0.053} & {61.484$\pm$0.037}\\
1.476--20 & {61.456$\pm$0.052} & {66.443$\pm$0.054} & {63.950$\pm$0.038}\\ 
0.9958-20 & {63.431$\pm$0.053} & {68.994$\pm$0.056} & {66.213$\pm$0.039}\\
0.818--20 & {63.907$\pm$0.053} & {70.016$\pm$0.057} & {66.961$\pm$0.039}\\   
0.723--20 & {64.093$\pm$0.053} & {70.536$\pm$0.057} & {67.314$\pm$0.039}\\
0.475--20 & {64.875$\pm$0.053} & {71.605$\pm$0.057} & {68.240$\pm$0.039}\\
0.281--20 & {67.563$\pm$0.054} & {72.646$\pm$0.057} & {70.105$\pm$0.039}\\
\hline
\end{tabular}
\label{tab1}
\end{table}

\begin{figure}
\resizebox{0.75\textwidth}{!}{

  \includegraphics{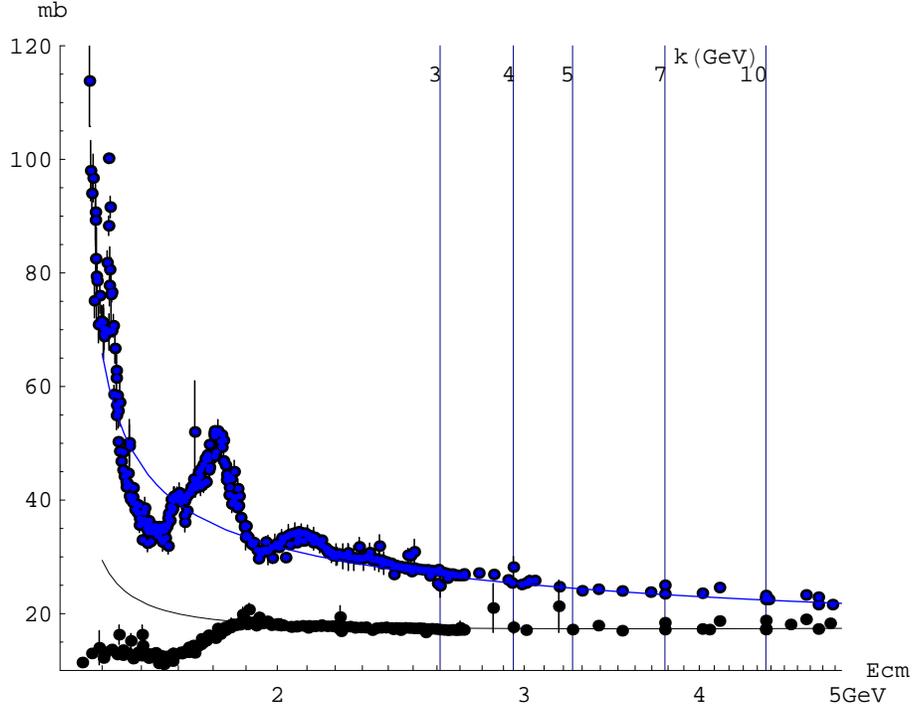}

}



\caption{The experimental $\sigma_{\rm tot}$ of $K^-p$(blue points) and 
of $K^+p$(black points) in the low-energy region: 
The vertical lines correspond to the laboratory momenta 
$k$=$\overline{N_1}$=10,7,5,4,3 GeV, which are 
selected as the lower limits of the FESR integrals (\ref{eq11}). 
Their numbers(in GeV) are shown in the upper side. 
The horizontal axis represents corrresponding center of mass energy $E_{cm}$.
The solid lines represent the low-energy extensions of our best fit using FESR 
in the case $\overline{N_1}=5$GeV.
}

\label{fig2}

\end{figure}

\begin{figure}
\resizebox{0.75\textwidth}{!}{

  \includegraphics{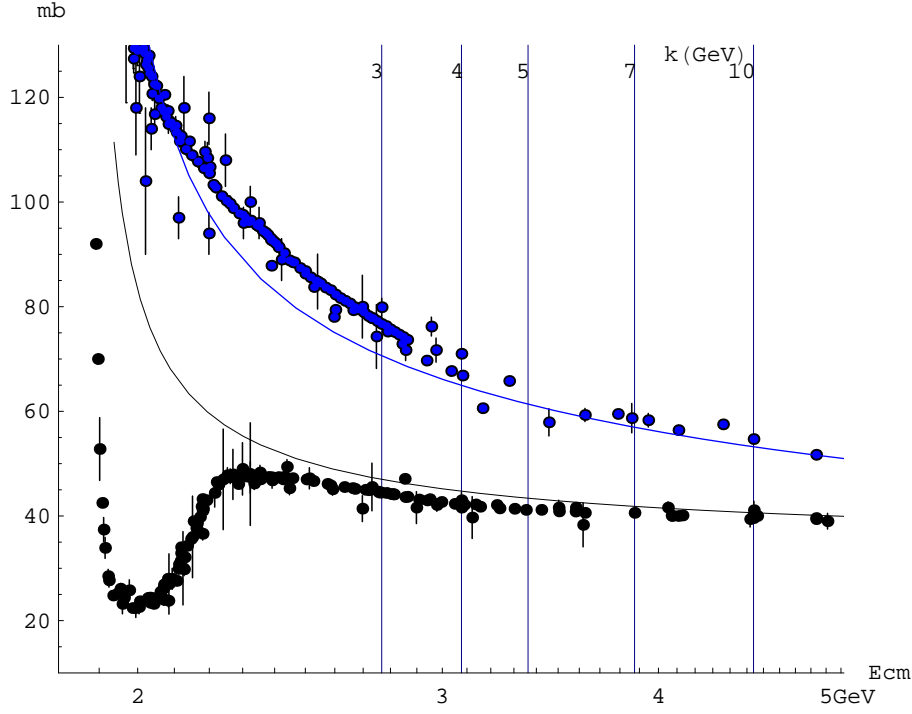}

}



\caption{The experimental $\sigma_{\rm tot}$ of $\bar pp$(blue points) and 
of $pp$(black points) in low-energy region: 
The vertical lines correspond to the values of $k=\overline{N_1}$
selected in our analyses. See, the caption in Fig.\ref{fig2}.
}

\label{fig3}

\end{figure}

Situations are different in $K^\mp p$ and $\bar pp,pp$ scatterings.
The $K^-p$ is the exothermic reaction.
$K^-p\rightarrow \pi^0\Lambda ,\pi\Sigma$ could occur even at threshold $\nu =m_K$, and 
the $\sigma_{\rm tot}^{K^-p}(k)$ increases like $1/k$ near threshold.
In the case of too small values of $\overline{N_1}$,
the integral of $\sigma_{\rm tot}$ is affected strongly by the contribution 
from these open channels.
Furthermore, in the exotic $K^+p$ channel there is a sudden decrease
of $\sigma_{\rm tot}$ observed below $E_{cm}\simeq 1.9$GeV($k\simeq 1.2$GeV).
Similarly, $\bar pp$ has a number of open meson channels below threshold, $\nu <M$,
and a big dip structure is observed in $\sigma_{\rm tot}$ of exotic ${pp}$ channel below 
$E_{cm}\simeq 2.2$GeV($k\simeq 1.4$GeV).  
We can see the situations in the $\sigma_{\rm tot}^{K^\mp p}$ and 
$\sigma_{\rm tot}^{\bar pp,pp}$ data shown in Figs.~\ref{fig2} and \ref{fig3}, 
respectively.
The reason for producing these strucures is not known, but it is safe to take
$\overline{N_1}$ to be fairly larger than $m_K$ and $M$.
Actually 
we take $\overline{N_1}\ge 3$GeV as 
$\overline{N_1}=10,7,5,4,3$GeV in $K^\mp p$ and $\bar pp,pp$ scatterings.
These laboratory momenta are represented by vertical lines in Fig.\ref{fig2} 
and \ref{fig3}, respectively.

For each value of $\overline{N_1}$
the integrals of $\sigma_{\rm tot}^{\bar ap,ap}$ and their average $\sigma_{\rm tot}^{(+)}$ 
are estimated by using the phenomenological fits to $K^\mp p$ and $\bar p(p)p$ scatterings.
The results are given in Tables \ref{tab2} and \ref{tab3}, 
respectively.

The values of integrals given in Tables \ref{tab1}, \ref{tab2} and \ref{tab3} 
are estimated with very small errors, which are generally less than 0.3\% .
We regard central values as exact ones and treat the FESR~(\ref{eq11}) 
as exact constraints between the fitting parameters.  

\begin{table}
\caption{The integral of $\sigma_{\rm tot}$ in $K^\mp p$:
$\frac{1}{2\pi^2}\int_{\overline{N_1}}^{\overline{N_2}}
\sigma_{\rm tot}^{K^-p,K^+p,(+)}(k)\ dk\ ({\rm GeV}^{-1})$.
The $\overline{N_2}$ is taken to be 20GeV while 
$\overline{N_1}$ are taken to be 10,7,5,4,3GeV, respectively.
}
\begin{tabular}{c|ccc}
$\overline{N_1}$--$\overline{N_2}$(GeV)
 & $\frac{1}{2\pi^2}\int_{\overline{N_1}}^{\overline{N_2}}
\sigma_{\rm tot}^{K^-p}(k)\ dk$
 & $\frac{1}{2\pi^2}\int_{\overline{N_1}}^{\overline{N_2}}
\sigma_{\rm tot}^{K^+p}(k)\ dk$
 & $\frac{1}{2\pi^2}\int_{\overline{N_1}}^{\overline{N_2}}
\sigma_{\rm tot}^{(+)}(k)\ dk$
  \\
\hline
10--20 & 28.217$\pm$0.068 & 22.661$\pm$0.053 & 25.439$\pm$0.043\\
 7--20 & 37.175$\pm$0.094 & 29.377$\pm$0.069 & 33.276$\pm$0.058\\
 5--20 & 43.425$\pm$0.110 & 33.818$\pm$0.077 & 38.622$\pm$0.067\\
 4--20 & 46.693$\pm$0.116 & 36.031$\pm$0.080 & 41.362$\pm$0.070\\
 3--20 & 50.120$\pm$0.120 & 38.257$\pm$0.081 & 44.189$\pm$0.072\\
\hline
\end{tabular}
\label{tab2}
\end{table}

\begin{table}
\caption{The integral of $\sigma_{\rm tot}$ in $\bar pp,pp$:
$\frac{1}{2\pi^2}\int_{\overline{N_1}}^{\overline{N_2}}
\sigma_{\rm tot}^{\bar pp,pp,(+)}(k)\ dk\ ({\rm GeV}^{-1})$.
The $\overline{N_2}$ is taken to be 20GeV, while 
$\overline{N_1}$ is taken to be 10,7,5,4,3GeV.
}
\begin{tabular}{c|ccc}
$\overline{N_1}$--$\overline{N_2}$(GeV)
 & $\frac{1}{2\pi^2}\int_{\overline{N_1}}^{\overline{N_2}}
\sigma_{\rm tot}^{\bar pp}(k)\ dk$
 & $\frac{1}{2\pi^2}\int_{\overline{N_1}}^{\overline{N_2}}
\sigma_{\rm tot}^{pp}(k)\ dk$
 & $\frac{1}{2\pi^2}\int_{\overline{N_1}}^{\overline{N_2}}
\sigma_{\rm tot}^{(+)}(k)\ dk$
  \\
\hline
10--20 &  65.75$\pm$0.24 & 51.07$\pm$0.06 &  {58.41$\pm$0.12}\\
 7--20 &  87.61$\pm$0.33 & 66.68$\pm$0.07 &  {77.14$\pm$0.17}\\
 5--20 & 103.48$\pm$0.39 & 77.28$\pm$0.08 &  {90.38$\pm$0.20}\\
 4--20 & 112.11$\pm$0.41 & 82.72$\pm$0.08 & {97.41$\pm$0.21}\\
 3--20 & 121.51$\pm$0.41 & 88.33$\pm$0.08 & {104.92$\pm$0.21}\\
\hline
\end{tabular}
\label{tab3}
\end{table}

\subsection{Analysis of $\pi^\mp p$ scattering}

The data\cite{r6} of $\sigma_{\rm tot}^{\pi^\mp p}$ for $k\ge 20$ GeV and 
$\rho^{\pi^\mp p}$ (more exactly Re~$f^{\pi^\mp p}$) for $k\ge 5$ GeV are 
fitted simultaneously.
In the FESR~(\ref{eq11}), $\overline{N_2}$ is taken to be 20GeV.
The $\overline{N_1}$ are chosen to be
10, 7, 5, 4, 3.02, 2.035, 1.476, 0.9958, 0.818, 0.723, 0.475, 0.281 GeV, 
as explained before.

For each value of $\overline{N_1}$, the integrals of $\sigma_{\rm tot}^{(+)}$
which are the RHS of Eq.~(\ref{eq11})
are given in Table \ref{tab1}.
The integral of asymptotic formula~(\ref{eq6}) which appears
in the LHS of Eq.~(\ref{eq11}) is calculable analytically, and we obtain 
the explicit form of FESR (\ref{eq11}). 
In the case of $\overline{N_1}=0.818$GeV, for example, the FESR (\ref{eq11}) is given by
\begin{eqnarray}
(\pi p) & \ \ &
 87.1714 \beta_{P^\prime} + 627.26 c_0 + 2572.37 c_1 + 10891.2 c_2
                  = 66.961 \pm 0.039\ .\ \ \ \ \ \ 
\label{eq14}
\end{eqnarray}
The error of the RHS is very small, and Eq.~(\ref{eq14}) is regarded as 
an exact constraint between the parameters, $c_{2,1,0}$ and $\beta_{P^\prime}$.
The $\beta_{P^\prime}$ is represented by the other three parameters $c_{2,1,0}$
as $\beta_{P^\prime}=\beta_{P^\prime}(c_2,c_1,c_0)$. 
The fitting is performed with five parameters, including $\beta_V$ and $F^{(+)}(0)$.
The $(\alpha_{P^\prime},\alpha_V)$ in Eqs.~(\ref{eq6}) and (\ref{eq7}) are fixed 
to be empirical values
$(0.500,0.497)$\cite{r15} in all the fitting procedures.
The values of $c_2$ and $\chi^2$ in the best fits for the respective $\overline{N_1}$ 
are given in Tables \ref{tab4}\footnote{
We adopt a different treatment of data from ref.\cite{r5},
where systematic errors were taken to be larger than those 
in the original ones\cite{r6}.}.

\begin{table}
\caption{
Values of $c_2$ of $\pi p$ scattering in the best fit with five-parameters, 
using FESR as a constraint.
The fitting $\chi^2$ are given in the next row.
The number of data points is 162.
The result of six-parameter fit 
without using FESR is also shown in the last column as No SR.
}
\begin{tabular}{c|ccccccc}
$\overline{N_1}$ & 10 & 7 & 5 & 4 & 3.02 & 2.035 & 1.476\\
\hline
$c_2(10^{-5})$ & 142(21) & 136(19) & 132(18) & 129(17) & 124(16) & 117(15) & 116(14)\\
${\chi^2_{\rm tot}}$ & ${149.05}$  & ${149.35}$
 & ${149.65}$  & ${149.93}$  & ${150.44}$
 & ${151.25}$  & ${151.38}$ \\  
\hline\hline
$\overline{N_1}$ & 0.9958 & 0.818 & 0.723 & 0.475 & 0.281 & & No SR\\
\hline
$c_2(10^{-5})$  & 116(14) & 121(13) & 126(13) & 140(13) & 121(12) & & 164(29)\\
${\chi^2_{\rm tot}}$  & ${151.30}$  & ${150.51}$
 & ${149.90}$ & ${148.61}$  & ${150.39}$
 &  & ${147.78}$ \\ 
\end{tabular}
\label{tab4}
\end{table}

As is seen in Table \ref{tab4}, values of the best fitted $c_2$
are almost independent of the choices of $\overline{N_1}$ 
(except for the case $\overline{N_1}$=0.475GeV).
%
The results are surprisingly stable, although  
there are many resonant strucures and $\sigma_{\rm tot}^{\pi^\mp p}$ 
show sharp peak and dip structures in this energy region. 

We can adopt the case of $\overline{N_1}$=0.818GeV as a representative 
of our results. The best-fit value of $c_2$ is
\begin{eqnarray}
c_2 &=& (121\pm 13)\cdot 10^{-5}\ \ .
\label{eq15}
\end{eqnarray}
The best-fit values of the other parameters are given in
Table~\ref{tab4a}.

\begin{table}
\caption{Values of best-fitted parameters 
and their one-standard deviations in $\pi^\mp p$ scattering.
The FESR with $\overline{N_1}$=0.818GeV is used and $\beta_{P^\prime}$
is obtained from the other parameters by using this FESR.
Our predicted lines in Fig. \ref{fig5} 
are depicted by using these values.
}
\begin{tabular}{cccccc}
 $c_2(10^{-5})\ $ & $c_1$ & $c_0$ & $F^{(+)}(0)\ $ & $\beta_V$
    & $\beta_{P^\prime}$\\
\hline
121.1 & -0.01179 & 0.1141 & -0.0180 & 0.04004 & 0.1437\\
\hline
$+13.3$ & -0.01385 & 0.1224 & -0.2785 & 0.03991 & 0.1280\\
$-13.3$ & -0.00974 & 0.1058 & 0.2425 & 0.04018 & 0.1594\\
\hline
\end{tabular}
\label{tab4a}
\end{table}

The result should be compared to the analysis with no use of FESR.
In this case there is no constraint between parameters, and the fitting
is performed with six parameters including $\beta_{P^\prime}$.
The value of $c_2$ in the best fit is
\begin{eqnarray}
c_2 &=& (164\pm 29)\cdot 10^{-5}\ \ .
\label{eq16}
\end{eqnarray}
This value should be compared with the result (\ref{eq15}) using FESR.
By including rich information of low-energy scattering data in the form of FESR,
the error of $c_2$ in Eq.~(\ref{eq15}) becomes less than half of Eq.~(\ref{eq16}).

\begin{figure}
\resizebox{0.75\textwidth}{!}{

  \includegraphics{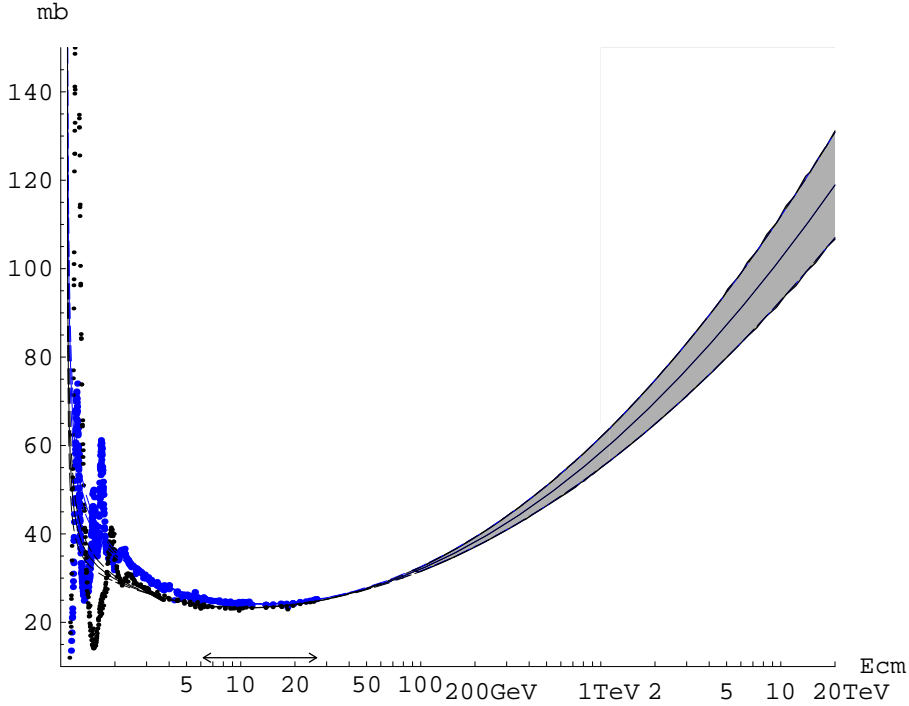}

}



\caption{Prediction of $\sigma_{\rm tot}^{\pi p}$ with No use of FESR
The data points are given with no error bars. 
The big blue points (line) are data (best-fitted curve) for $\pi^- p$.
The black points and lines are for $\pi^+ p$. The horizontal arrow represents the energy region
of the fitting.
The shaded region corresponds to the uncertainty
of the prediction by the best fit, where $c_2=(164\pm 29)\times 10^{-5}$. 
The $c_2$ has large uncertainty since it is not determined well 
by direct fitting of the data above $k=20$GeV($E_{cm}=6.2$GeV).
}

\label{fig4}

\end{figure}

\begin{figure}
\resizebox{0.75\textwidth}{!}{

  \includegraphics{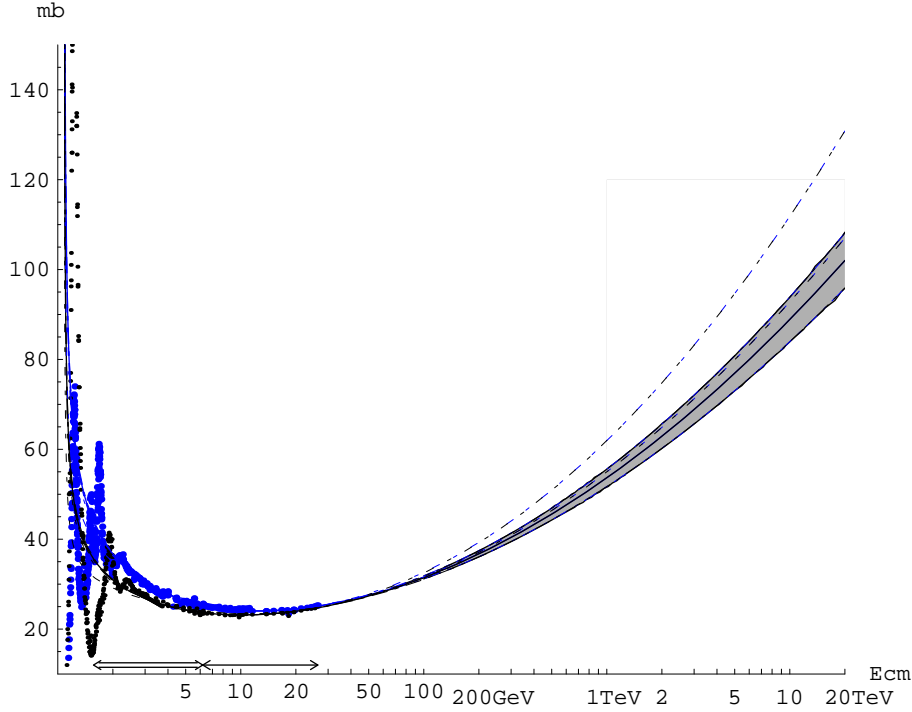}

}


\caption{Prediction of $\sigma_{\rm tot}^{\pi p}$ with the use of FESR
 in the case $\overline{N_1}=0.818$GeV as a constraint.
The data points are given with no error bars. 
The big blue points (line) are data (best-fitted curve) for $\pi^- p$.
The black points and lines are for $\pi^+ p$. Single horizontal arrow represents the energy region
of the fitting, while double horizontal arrow represents
the energy region of the FESR integral, 
$k=\overline{N_1}$ through $\overline{N_2}$(=20GeV).
The uncertainty of the prediction by the best fit is shown by shaded region, 
where $c_2=(121\pm 13)\times 10^{-5}$. It is greatly improved from that with 
no use of FESR, shown by dashed line.
The inclusion of rich information of low-energy scattering data through FESR is 
essential to determine $c_2$. 
%
}

\label{fig5}

\end{figure}

Predicted spectra of $\sigma_{\rm tot}$ with no use of FESR are shown in Fig.~\ref{fig4} 
and with using FESR in Fig.~\ref{fig5}.
The uncertainties of the prediction by the best fit are 
shown by shaded region in Figs. \ref{fig4} and \ref{fig5}, 
which corresponds to $c_2$ in Eqs.~(\ref{eq16}) and (\ref{eq15}), respectively.
Our prediction in Fig.\ref{fig5} is greatly improved from that in Fig.~\ref{fig4}.

\begin{figure}
\resizebox{0.5\textwidth}{!}{

  \includegraphics{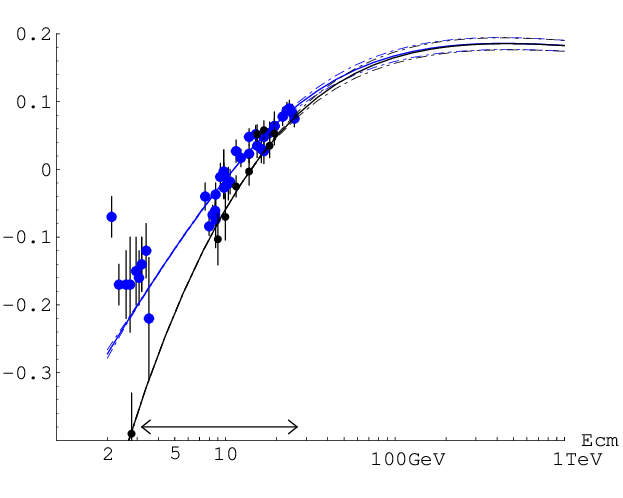}

}



\caption{Result of the fit and prediction of $\rho^{\pi^\mp p}$ with the use of FESR
 in the case $\overline{N_1}=0.818$GeV as a constraint.
The big blue points (line) are data (best-fitted curve) for $\pi^- p$.
The black points and lines are for $\pi^+ p$. 
A horizontal arrow represents the energy region
of the fitting. 
}

\label{fig5a}

\end{figure}

The result of the fit to $\rho^{\pi^\mp p}$ are given in Fig.~\ref{fig5a}.

The $c_2$log$^2\frac{\nu}{\mu}+c_1$log~$\frac{\nu}{\mu}$ with $c_2>0$ shows 
the shape of parabola as a function of log~$\frac{\nu}{\mu}$ with minimum.
In order to determine the $c_2,c_1$ coefficients of log$^2\frac{\nu}{\mu}$ and 
log~$\frac{\nu}{\mu}$
with a sufficient accuracy, a few orders of magnitude are neccesary 
for fitted energy region.
The fitted energy region for $\pi^- p$ is 20GeV$<k<370$GeV for $\pi^-p$
shown by horizontal arrow in the figure, which 
corresponds to 6.2GeV$<E_{cm}(\equiv\sqrt s)<$26.4GeV.
This energy range is insufficient to determine $c_2$ with enough accuracy.

The energy region of the FESR integral, $\overline{N_1}$(=0.818GeV)$<k<$20GeV 
(which corresponds to $1.56<\sqrt s=E_{cm}<6.2$GeV), 
is shown by double horizontal arrow in Fig.~\ref{fig5}.
Additional information from this energy region greatly helps to improve our estimate
of $c_2$.
It is very important to include the information of the data 
in low-energy region by using FESR.

The value of $\sigma_{\rm tot}$ in $\pi^- N$ scattering at $k=610$GeV($E_{cm}=33.8$GeV) 
is reported by SELEX collaboration\cite{r14} as $\sigma_{\rm tot}^{\pi N}=26.6\pm 0.9$mb.
Here $N$ is not identified with proton or neutron.
Our prediction of $\sigma_{\rm tot}^{(+)}
(=(\sigma_{\rm tot}^{\pi^-p}+\sigma_{\rm tot}^{\pi^+p})/2 )$ at this energy is 
$25.75\pm 0.05$mb,
and $\sigma_{\rm tot}^{\pi^-p}-\sigma_{\rm tot}^{\pi^+p}$ is 0.30mb.
By taking this value into account we may predict 
$\sigma_{\rm tot}^{\pi^- N}=25.8\pm 0.3$mb at $k$=610GeV. 
It is consistent with the SELEX measurement. 

In Fig.~\ref{fig1},
our best fitted curve in the case of $\overline{N_1}=0.818$GeV is also depicted.
The fitted energy region is above $k=20$GeV($E_{cm}=6.2$GeV), and 
it is far above the energy region shown in this Fig.\ref{fig1}.
Nevertheless, the low-energy extensions of the asymptotic formula  
almost coincide with the experimental $\sigma_{\rm tot}^{\pi^\mp p}$ 
in $E_{cm}$ up to $\sim 3$GeV, and in $E_{cm}< \sim 3$GeV
they seem to cross the averages of peak and dip structures
of various $N$ and $\Delta$ resonances. 
This shows that the FESR duality is satisfied in our best fit.

\subsection{Analysis of $K^\mp p$ scattering}

The $K^-p$ and $K^+p$ scatterings are analyzed 
by the same method.
We fix $\overline{N_2}=20$GeV while we take various values of $\overline{N_1}$
as $\overline{N_1}=10,7,5,4,3$GeV. 
The integral of $\sigma_{\rm tot}^{(+)}$
$(=(\sigma_{\rm tot}^{K^-p}+\sigma_{\rm tot}^{K^+p})/2)$
for each value of $\overline{N_1}$ is given in Table \ref{tab2}, 
and the FESR~(\ref{eq11}) is written in an explicit form.
In the case of $\overline{N_1}=5$GeV, the FESR is given by
\begin{eqnarray}
(Kp) &\ \ \  & 8.21363 \beta_{P^\prime} + 39.2291 c_0 + 124.142 c_1 + 398.549 c_2
                = 38.62 \pm 0.07\ ,\ \ \ \ \ \ \ 
\label{eq17}
\end{eqnarray}
which is regarded as a constraint between parameters, $c_{2,1,0}$ and $\beta_{P^\prime}$.
Solving this constraint, the $\beta_{P^\prime}$ is represented by the other three parameters. 
The experimental $\sigma_{\rm tot}^{K^\mp p}$ in $k\ge 20$GeV and 
$\rho^{K^\mp p}$(more exactly Re $f^{K^\mp p}$) in $k\ge 5$GeV are fitted simultaneously
with five parameters, $c_{2,1,0},\beta_{V}$ and $F^{(+)}(0)$, 
where $\alpha_{P^\prime},\alpha_V$ are fixed with common values to $\pi p$ case, 
$\alpha_{P^\prime}=0.5,\alpha_V=0.497$. 
The fits are successful, independently of the choices of $\overline{N_1}$,
as shown in Table \ref{tab5} where we only show the values of $c_2$ and the total $\chi^2$
in the best fits.

\begin{table}
\caption{Values of $c_2$ and the $\chi^2$ in the best fits 
to $\sigma_{\rm tot}^{K^\mp p}$ and $\rho^{K^\mp p}$. 
The FESR with the integral $\overline{N_1}$ through $\overline{N_2}$ is used as 
constraints. The $\overline{N_2}$ is fixed to 20GeV while we take various values of 
$\overline{N_1}$. 
The number of data points is 111, fitted with 5 parameters in the case with using FESR.
The result with no use of FESR is given in the last column as No SR.}
\begin{tabular}{c|ccccc|c}
$\overline{N_1}$(GeV) & 10 & 7 & 5 & 4 & 3 & No SR \\  
\hline
$c_2(10^{-4})$ & 179(61) & 176(54) & 176(49) & 176(47) & 174(44) & 266(95) \\ 
$\chi^2_{\rm tot}$ & 64.01 & 63.90 & 63.80 & 63.76 & 63.77 & 62.29\\
\end{tabular}
\label{tab5}
\end{table}

\begin{table}
\caption{Values of best-fitted parameters 
and their one-standard deviations in $K^\mp p$ scattering.
The FESR with $\overline{N_1}$=5GeV is used and $\beta_{P^\prime}$
is obtained from the other parameters by using this FESR.
Our predicted lines in Fig. \ref{fig7} 
are depicted by using these values. }
\begin{tabular}{cccccc}
$c_2(10^{-4})$ & $c_1$ & $c_0$
 & $F^{(+)}(0)$ & $\beta_V$ & $\beta_{P^\prime}$\\ 
\hline
 175.7 & -0.1388 & 1.207 & 1.660 & 0.5684 & 0.1840\\
\hline
 +49.5 & -0.2042 & 1.439 & 0.640 & 0.5668 & -0.1775\\
 -49.5 & -0.0733 & 0.974 & 2.680 & 0.5699 &  0.5455\\
\hline
\end{tabular}
\label{tab6}
\end{table}

The central values of $c_2$ in the best fits are very stable, and almost independent
of the choices of $\overline{N_1}$. We choose $\overline{N_1}=5$GeV as a representative
of our results. This value is fairly larger than the momentum of the dip structure 
observed in $\sigma_{\rm tot}^{K^+p}$ below $k\simeq 1.2$GeV.  
The best-fitted value of $c_2$ is 
\begin{eqnarray}
c_2 &=& (176\pm 49)\cdot 10^{-4}\ .
\label{eq18}
\end{eqnarray}
The values of the other parameters are given in Table \ref{tab6}.
The results are compared with the analysis with no use of FESR, 
\begin{eqnarray}
c_2 &=& (266\pm 95)\cdot 10^{-4}\ .
\label{eq19}
\end{eqnarray}
This uncertainty of $c_2$ is very large, since
the range of momenta of the fitting 20GeV$<k<$310GeV, which corresponds to 
6.2GeV$<E_{cm}<$24.1GeV, is insufficient to determine $c_2$
accurately by direct fitting.
The value of $\beta_{P^\prime}$ in the best fit
becomes negative and unphysical. Thus,  
the central value $c_2=266\times 10^{-4}$ is considered to be too large and unreliable.  

\begin{figure}
\resizebox{0.75\textwidth}{!}{

  \includegraphics{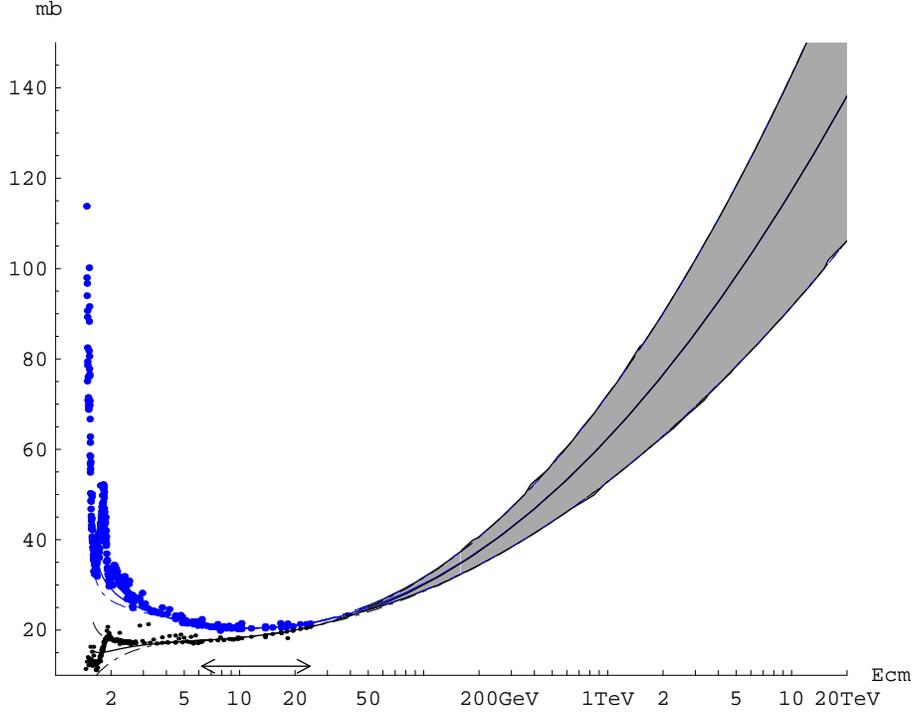}

}



\caption{Prediction of $\sigma_{\rm tot}^{K p}$ with No use of FESR in 
The data points are given with no error bars. 
The big blue points (line) are data (best-fitted curve) for $K^- p$.
The black points and lines are for $K^+ p$. The horizontal arrow represents the energy region
of the fitted data.
The shaded region represents the uncertainty
of the prediction, where $c_2=(266\pm 95)\times 10^{-4}$. 
The $c_2$ has a very large uncertainty since it is not determined well 
by direct fitting to the high-energy experimental data.
}

\label{fig6}

\end{figure}

\begin{figure}
\resizebox{0.75\textwidth}{!}{

  \includegraphics{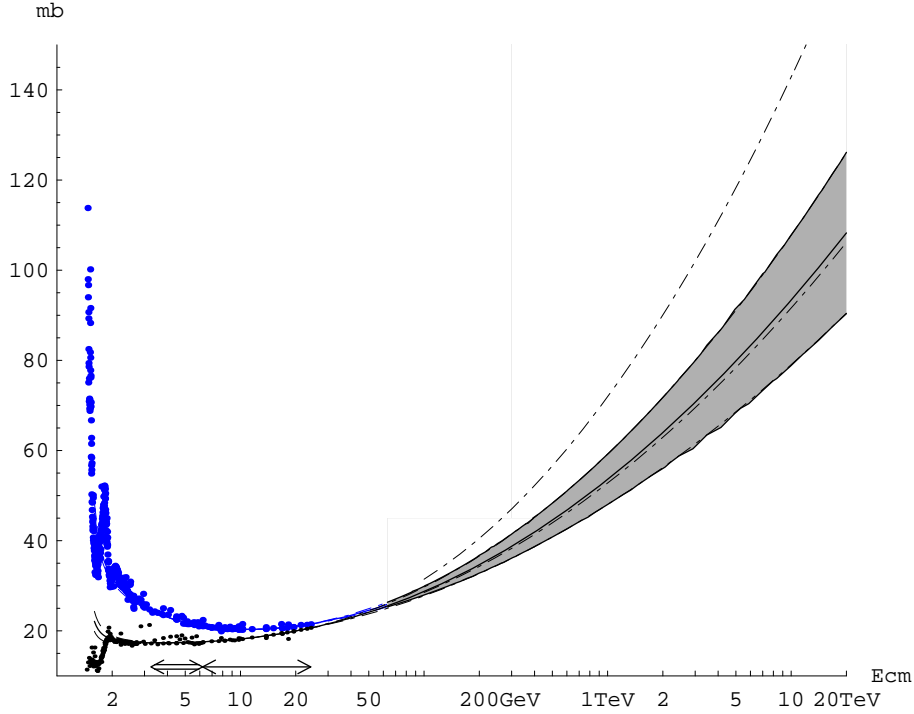}

}



\caption{Prediction of $\sigma_{\rm tot}^{Kp}$ with using FESR 
in the case $\overline{N_1}=5$GeV as a constraint:
The big blue points (line) are data (best-fitted curve) for $K^- p$.
The black points and lines are for $K^+ p$. Single horizontal arrow represents the energy region
of the fitted data, while  double horizontal arrow represents the energy region of 
the FESR integral, $k=\overline{N_1}$ through $\overline{N_2}$(=20GeV).
The data points are given with no error bars. 
The uncertainty of the prediction by the best fit, 
shown by shaded region, is improved from that with no use of FESR, represented by
dot-dashed line. 
The best-fitted $c_2$ is $c_2=(176\pm 49)\times 10^{-4}$.
The inclusion of the information of low-energy data by FESR is essential to 
improve the estimation of $c_2$.
}

\label{fig7}

\end{figure}

\begin{figure}
\resizebox{0.5\textwidth}{!}{

  \includegraphics{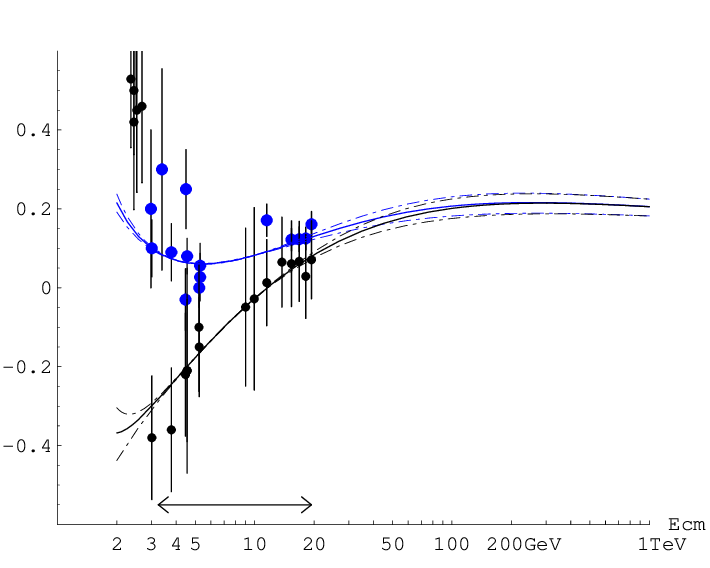}

}



\caption{Result of the fit and prediction of $\rho^{K^\mp p}$ with the use of FESR
 in the case $\overline{N_1}=5$GeV as a constraint.
The big blue points (line) are data (best-fitted curve) for $K^- p$.
The black points and lines are for $K^+ p$. 
A horizontal arrow represents the energy region
of the fitting. 
}

\label{fig7a}

\end{figure}

By including the data above $k=5$GeV($E_{cm}=3.25$GeV) in the form of FESR,
the error of $c_2$ in Eq.~(\ref{eq18}) becomes about half
of Eq.~(\ref{eq19}).
Correspondingly, the predicted spectra with using FESR given in Fig.~\ref{fig7} 
are greatly improved from Fig.~\ref{fig6} with no use of FESR.
The inclusion of the information in low-energy region is essential
to determine the value of $c_2$ reliably in $Kp$ scattering.
The fitting results and prediction of $rho^{K^-p,K^+p}$ using FESR are given in Fig.~\ref{fig7a}.

The figure \ref{fig2} shows the data in the low-energy region. 
The low-energy extensions of our best fitted curves in the case of $\overline{N_1}$=5GeV
are also depicted in this figure.
They reproduce surprisingly well the experimental $\sigma_{\rm tot}^{K^-p}$ 
and $\sigma_{\rm tot}^{K^+p}$, although the energy region of the FESR integral and 
the energy region of the fitting are 
above the energy region shown in this figure.
This shows that our best fitted curves satisfy the FESR duality.

\subsection{Analysis of $\bar pp,pp$ scatterings}

Similarly to the $Kp$ scatterings, in the analysis of $\bar pp,pp$ scatterings 
we fix $\overline{N_2}=20$GeV, while we take various values of $\overline{N_1}$
as $\overline{N_1}=10,7,5,4,3$GeV.
For each value of $\overline{N_1}$, the FESR~(\ref{eq11}) is written in an 
explicit form, where the integral of $\sigma_{\rm tot}^{(+)}$
$(=(\sigma_{\rm tot}^{\bar pp}+\sigma_{\rm tot}^{pp})/2)$ is given in
Table \ref{tab3}. 
In the case $\overline{N_1}=5$GeV, for example, the FESR is given by
\begin{eqnarray}
 & (pp)\ \ \  &\ 3.14058 \beta_{P^\prime} + 10.8947 c_0 + 27.5046 c_1 + 71.0017 c_2
              = 90.38 \pm 0.20\  , \ \ \ \ \ \ \ 
\label{eq20}
\end{eqnarray}
which is regarded as a constraint between parameters, $c_{2,1,0}$ and $\beta_{P^\prime}$,
and leads to the relation $\beta_{P^\prime}=\beta_{P^\prime}(c_2,c_1,c_0)$. 

The $\sigma_{\rm tot}^{\bar pp}$ data are obtained 
up to $E_{cm}=63$GeV by ISR experiment, up to $E_{cm}=0.9$TeV by SPS experiment,
and up to $E_{cm}=1.8$TeV by Tevatron experiment. 
There are two conflicting measurements in Tevatron experiment, 
by D0\cite{rE710,rE811} and CDF\cite{rCDF}. 
The very high-enegy data with large uncertainties are obtained 
by cosmic ray experiments.

The experimental $\sigma_{\rm tot}^{\bar pp,pp}$ in $k\ge 20$GeV and $\rho^{\bar pp,pp}$
in $k\ge 5$GeV, up to Tevatron energy, 
are fitted simultaneously.
The fittings are performed with five parameters, $c_{2,1,0},\beta_{V}$ and $F^{(+)}(0)$,
by using FESR as a constraint\footnote{
The original data of $\rho^{pp}$ gives no successful fits, since the 
data are mutually inconsistent with different experiments. 
Data points by refs.\cite{rFaj} and \cite{rBel} have comparatively small errors,
 and seem to be inconsistent with the other data sets by inspection.
We have tried to fit the data set only including \cite{rFaj} for $\rho^{pp}$
in the relevant energy region, but it is not successful.
We remove these two data sets from our best fit.
}.
The best fitted values of $c_2$ and total $\chi^2$
for the respective values of $\overline{N_1}$ are given in Table \ref{tab7}.
The result is compared with the six-parameter fit of the analysis without using FESR, 
denoted as No SR in the same table.
By considering this result, 
we choose $\overline{N_1}=5$GeV as a representative of our analyses.
The best-fitted $c_2$ is given by
\begin{eqnarray}
c_2 &=& (504\pm 26)\cdot 10^{-4}\ .
\label{eq21}
\end{eqnarray}
The values of the other parameters are given in Table \ref{tab7a}.

%
\begin{table}
\caption{
Values of $c_2$ and the total $\chi^2$ in the best fit 
to $\bar pp,pp$ scatterings, using FESR as a constraint. 
The data up to Tevatron energy, $E_{cm}=1.8$TeV, are fitted simultaneously.
The number of data points is 234, fitted by 5 parameters in the case using FESR as a constraint.
The result of six-parameter fit 
without using FESR is also shown in the last column as No SR.
}
\begin{tabular}{c|ccccc|c}
$\overline{N_1}$ & 10 & 7 & 5 & 4 & 3 & No SR\\
\hline
$c_2(10^{-4})$ & 505(28) & 506(27) & 504(26) & 500(26) & 493(25) & 491(34)\\
${\chi^2_{\rm tot}}$ & ${214.52}$  & ${214.53}$
 & ${214.32}$  & ${214.14}$  & ${213.99}$
 & ${213.98}$ \\  
\hline
\end{tabular}
\label{tab7}
\end{table}

\begin{table}
\caption{Values of best-fitted parameters 
and their one-standard deviations in $\bar pp,pp$ scattering.
The FESR with $\overline{N_1}$=5GeV is used and $\beta_{P^\prime}$
is obtained from the other parameters by using this FESR.
Our predicted lines in Figs. \ref{fig8} and \ref{fig8a} 
are depicted by using these values. }
\begin{tabular}{cccccc}
$c_2(10^{-4})$ & $c_1$ & $c_0$
 & $F^{(+)}(0)$ & $\beta_V$ & $\beta_{P^\prime}$\\ 
\hline
 503.6 & -0.2432 & 6.647 & 10.51 & 3.721 & 6.713\\
\hline
 $+26.1$ & -0.2810 & 6.788 & 10.24 & 3.728 & 6.495\\
 $-26.1$ & -0.2054 & 6.505 & 10.77 & 3.714 & 6.931\\
\hline
\end{tabular}
\label{tab7a}
\end{table}

The improvement from the result $c_2=(491\pm 34)\cdot 10^{-4}$ obtained
by the fit without using FESR 
is not large, since the high-energy data by SPS and Tevatron experiments,
which affects directly the estimation of $c_2$-value, are included in both fits.

\begin{figure}
\resizebox{0.75\textwidth}{!}{

  \includegraphics{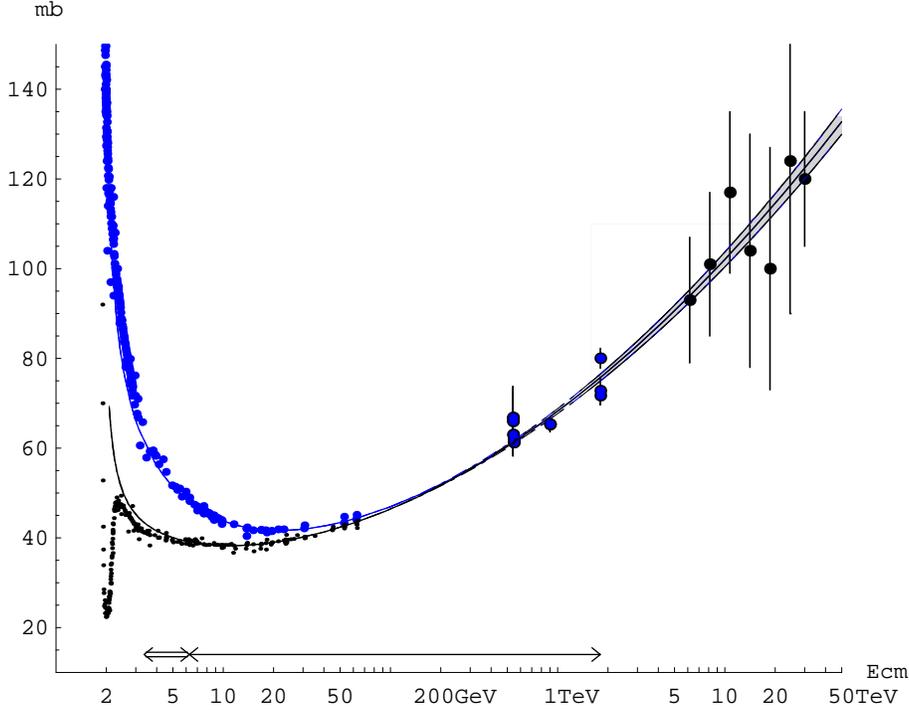}

}



\caption{Prediction of $\sigma_{\rm tot}^{\bar pp,pp}$ with using FESR:
The data up to Tevatron energy are fitted simultaneously.
The single horizontal arrow represents the energy region
of the fitting. 
The big blue points (line) are data (best-fitted curve) for $\bar pp$.
The black points and lines are for $pp$. The data points are given with no error bars. 
The double horizontal arrow represents the energy region
of the FESR integral, $k$=$\overline{N_1}$(=5GeV in this case) 
through $\overline{N_2}$(=20GeV).
The shaded region corresponds to the uncertainty
of our prediction in the best fit, where $c_2=(504\pm 26)\times 10^{-4}$.
}

\label{fig8}

\end{figure}

\begin{figure}
\resizebox{0.5\textwidth}{!}{

  \includegraphics{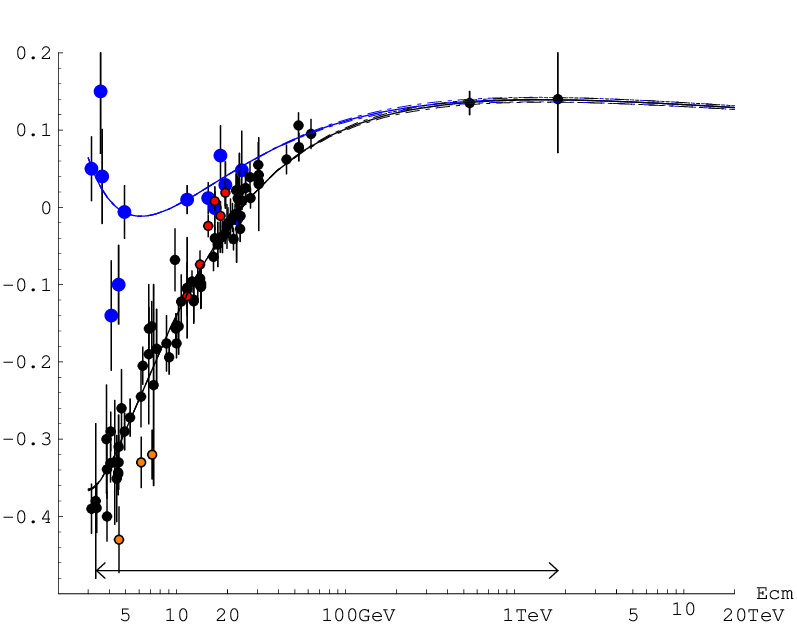}

}

\caption{Result of the fit and prediction of $\rho^{\bar pp,pp}$ with the use of FESR
 in the case $\overline{N_1}=5$GeV as a constraint.
The big blue points (line) are data (best-fitted curve) for $\bar pp$.
The black points and lines are for $pp$. A horizontal arrow represents the energy region
of the fitting. $\rho^{pp}$ data by ref.\cite{rFaj}(red points) and 
by ref.\cite{rBel}(orange points) are removed from our best fit.
}

\label{fig8a}

\end{figure}

Our predicted spectra of $\sigma_{\rm tot}^{\bar pp,pp}$ 
in the case of $\overline{N_1}$=5GeV are given in Fig.~\ref{fig8}.
Fitting result and prediction of $\rho^{\bar pp,pp}$ are given in Fig.~\ref{fig8a}.

\subsection{Test of the universality of $B$}

Using the values of parameters given in the previous subsections, 
we can test the universality of $B$ parameters from the experimental data
of $\pi^\mp p,\ K^\mp p$ and $\bar p(p)p$ scatterings. 

In ref.\cite{r6} the asymptotic formula of total cross section is represented 
in terms of squared CM energy $s$ in the form
\begin{eqnarray}
\sigma_{\rm tot}^{(+)} & \simeq & 
Z^{ap}+ B~{\rm log}^2\frac{s}{s_0} \ ,  
\label{eq22}
\end{eqnarray}
which is already given in Eq.~(\ref{eq3}).
In ref.\cite{r6} $B$ and $s_0$ is assumed to be universal in the relevant 
processes while $Z_{ap}$ are taken to be 
process-dependent\footnote{
Ref.\cite{r6} treat the scatterings of $\bar p(p)p,\bar p(p)n,\Sigma^-p,\pi^\mp p,
K^\mp p,K^\mp n,\gamma p,\gamma\gamma$.
All data sets are fitted by using a common value of $B$ (and a common value of $s_0$). 
The resulting $B$ is $B=0.308(10)$mb.
}.

We use the asymptotic formula of crossing-even amplitude~(\ref{eq6}) 
which gives
\begin{eqnarray}
\sigma_{\rm tot}^{(+)} & \simeq & \frac{4\pi}{m^2}\left(
c_0 + c_1 {\rm log}\frac{\nu}{m} + c_2 {\rm log}^2\frac{\nu}{m} \right)\ .  
\label{eq23}
\end{eqnarray}
It is the same equation as Eq.~(\ref{eq2}).
Here we omit the $P^\prime$-term proportional to $\beta_{P^\prime}$ and
 use the approximation $\nu /k \simeq 1$ in high energies.

Using the relation $s \simeq 2M\nu$ at high-energies from Eq.~(\ref{eq1}), 
we obtain the relation between the  parameters
\begin{eqnarray}
B &=& \frac{4 \pi}{m^2} c_2\ ,\ \  \label{eq24}\\
Z^{ap} &=& \frac{4 \pi}{m^2}\left( c_0 -\frac{c_1{}^2}{4 c_2} \right)\ ,\ \ \label{eq25}\\ 
s_0 &=& 2 M m\ {\rm exp}\left[ -\frac{c_1}{2 c_2} \right] + M^2 + m^2  \ \ .
\label{eq26}
\end{eqnarray}
The parameters $c_{2,1,0}$ are treated as process-dependent in our analyses.
Substituting the best-fitted values of $c_{2,1,0}$ into these equations,
we can estimate the values of $B,Z^{ap},\sqrt{s_0}$ individually
for $\pi p,Kp$ and $\bar pp,pp$ scatterings without using the universality hypothesis. 
The results are given in Table \ref{tab8}.
\begin{table}
\caption{Values of $B,\ Z^{ap}$ and $\sqrt s_0$ obtained from our best fits
with using FESR(LHS) and without using FESR (RHS).
The $B,\ Z^{ap}$ and $\sqrt s_0$ are estimated individually for
the processes of $\pi p,\ Kp$ and $\bar p(p)p$ scatterings.
}
\begin{tabular}{c|ccc||ccc}
 & $B$(mb) & $Z^{ap}$(mb) & $\sqrt{s_0}$(GeV)
  & $B$(mb) & $Z^{ap}$(mb) & $\sqrt{s_0}$(GeV)\\
\hline
$\pi p$ & 0.304(34) & 21.45(32) & 5.92(90) & 0.411(73) & 22.99 & 9.71\\
$Kp$ & 0.354(99) & 18.7(1.1) & 7.1(2.6) & 0.535(190)  &20.67 & 12.14\\
$\bar p(p)p$ & 0.280(15) & 35.31(36) & 4.63(53) & 0.273(19) & 34.98 & 4.28\\
\end{tabular}
\label{tab8}
\end{table}
The results with using FESR are given in the LHS, and 
those without using FESR are given in the RHS.

As shown in this table, in the case of using FESR,
the values of $B$ for $\pi p,Kp$ and $\bar p(p)p$ scatterings 
denoted, respectively, as $B_{\pi p},B_{Kp}$ and $B_{pp}$, are mutually
consistent within one standard deviation.

In contrast, if we do not use the FESR as constraints, 
$B_{\pi p}$ and $B_{Kp}$ have large uncertainties, and we 
cannot obtain any definite conclusion.

It is very interesting that by including rich informations of low-energy
scattering data through FESR, the central values of $B$ become mutually closer, 
and consistent with each other.
The FESR duality suggests the universality of $B$. 

Another interesting feature is the values of $Z^{ap}$. 
If we neglect relatively small contributions from $\beta_{P^\prime}$ and $\beta_V$ terms
the $Z^{ap}$ are equal to the $\sigma_{\rm tot}$ at energy $\sqrt{s_0}$,
which is the lowest point of parabola (\ref{eq23}) of log$\frac{\nu}{m}$.
The values of $Z^{ap}$ for $\pi p,Kp$ and $\bar p(p)p$ approximately
satisfy the relation of the quark model $Z^{\pi p}:Z^{Kp}:Z^{pp}\simeq 2:2:3$. 
$Z^{ap}$ and the parameters $\beta_{P^\prime}$ and $\beta_V$
appearing in the Regge theory are controlled by non-perturbative soft physics of QCD,
while the term $B$~log$^2s/s_0$ describing the rise of $\sigma_{\rm tot}$
is plausibly universal for hadron-hadron scatterings. 

It is to be noted that 
this picture is inferred in an early work\cite{r9} where it is stated that
"the first, constant term[corresponding to $Z^{ap}$ in Eq.~(\ref{eq22}) 
in the present article] ... 
corresponds to a process-dependent(valence quark scattering)
contribution, whereas the second, logarithmically rising one is universal 
(gluon scattering)." "We conclude that it is the FNAL(-ISR) energy interval where
$\sigma^{MB}/\sigma^{BB}$(the ratio of the $\sigma_{\rm tot}$ for meson-baryon
scattering to the baryon-baryon scattering) comes 
closest to $2/3$. With the further increase of energy the ratio will approach unity."

In the arguments of color glass condensate(CGC) of QCD,
the gluon component of the target particle drastically increases
in high-energy scattering. This is based on the calculation of perturbative QCD.
The radius $R$ of the "black" region 
where strong absorption of incident particles occurs
increases plausibly by a log~$s$ term with a universal coefficient.
The $\sigma_{\rm tot}$ may be given by $2\pi R^2$ and 
 the factor $B$ is expected to be
universal for $\pi p,Kp$ and $\bar pp,pp$.

In ref.\cite{r6} the $s_0$ as well as $B$ are taken to be universal in the fittings. 
The $\sqrt{s_0}$ is not suggested to be process-independent 
both in ref.\cite{r9} and in the framework of CGC. 
In Table \ref{tab8} the values of $\sqrt{s_0}$ for the relevant three processes
seem to be closer to each other in the case of using FESR, compared with the case of no use of FESR. 
This possibility will be investigated in the next subsection.

\subsection{Analysis with common value of $B$}

Our analyses in previous subsections suggest the universality of $B$ 
in $\pi p,\ Kp$ and $\bar p(p)p$ scatterings. 
Now let us try to fit all the data by taking the same value of $B$ from the beginning.
The $\sigma_{\rm tot}$ above $k\ge 20$GeV and $\rho$ above $k\ge 5$GeV 
for $\pi^\mp p,K^\mp p,\bar p(p)p$ scatterings
are fitted simultaneously. There are three sets of parameters, 
$c_2,c_1,c_0,F^{(+)}(0),\beta_{P^\prime},\beta_V$. The three FESR 
(\ref{eq14}),(\ref{eq17}) and (\ref{eq20}) are used as constraints. 
Now three $c_2$ are not independent. They are represented 
by one universal $B$ parameter through Eq.~(\ref{eq24}).
So the number of fitting parameters is 13.
Successful fits are obtained with
the total  
$\chi^2 =429.55  = 150.83(\pi^\mp p) + 64.28(K^\mp p)+ 214.44(\bar p(p)p)$
with $(N_D-N_P)=(508-13)$. It is compared with the best fitted $\chi^2$ 
for respective data sets with no universality constraints, which are given in the 
tables \ref{tab4}, \ref{tab5} and \ref{tab7}:
$\chi^2/(N_D-N_P)=150.51/(162-5)$ for $\pi^\mp p$, $63.80/(111-5)$ for $K^\mp p$,
$214.32/(235-5)$ for $\bar p(p)p$. Their sum is 428.62.
The increase of total $\chi^2$ is, thus, only 0.93.
The constraint of universal $B$ is consistent with the present experimental data.

The value of $B$ in this universality fit is given by
\begin{eqnarray}
B &=& 0.285 \pm 0.013\ {\rm mb}.
\label{eq27}
\end{eqnarray}
The values of $\sqrt{s_0}$ and $Z^{ap}$ are also given in the upper half of Table \ref{tab9}.  

\begin{table}
\caption{Values of parameter in the best fits 
by assuming universality of $B$ and of $B$ and $s_0$,
which are taken to be common in fitting the data of relevant processes.
The $B$, $Z^{ap}$ and $\sqrt s_0$ are related with $c_2,c_0$ and $c_1$ 
by Eqs.~(\ref{eq24}), (\ref{eq25}) and (\ref{eq26}).
The $\beta_{P^\prime}$ are obtained from values of 
the other parameters through FESR. 
}
\begin{tabular}{c|cccccc}
 & $B$(mb) & $\sqrt{s_0}$(GeV) & $Z^{ap}$(mb)
   & $F^{(+)}(0)$ & $\beta_V$ & $\beta_{P^\prime}$\\
\hline
$\pi p$ & 0.285(13) & 5.40(37) & 21.24(16) & 0.13(61) & 0.04012(95) & 0.1527(62)\\
$Kp$ & 0.285(13) & 5.17(38) & 17.91(19) & 2.33(1.01) & 0.5618(82) & 0.4296(481)\\
$\bar p(p)p$ & 0.285(13) & 4.82(49) & 35.44(32) & 10.41(60) & 3.723(36) & 6.637(198)\\
\hline
$\pi p$ & 0.304(10) & 5.75(34) & 21.36(15) & -0.10(61) & 0.04043(93) & 0.1472(61)\\
$Kp$ & 0.304(10) & 5.75(34) & 18.19(16) & 2.11(1.01) & 0.5613(81) & 0.3535(461)\\
$\bar p(p)p$ & 0.304(10) & 5.75(34) & 36.04(17) & 9.88(53) & 3.745(35) & 6.232(122)\\
\hline
\end{tabular}
\label{tab9}
\end{table}

By taking the value of $B$ to be universal, 
the best-fitted values of $\sqrt s_0$
become closer to each other. 
Successful fits are obtained in ref.\cite{r6} by taking both $B$ and $s_0$ being
universal. We also try to fit the data by taking both $B$ and $s_0$
being common in the relevant three processes.
The resulting $\chi^2=435.24=152.19(\pi^\mp p)+64.06(K^\mp p)
+219.00(\bar p(p)p)$ with 11 parameters, and the fit is successful.
The increase of $\chi^2$ is totally 6.62 from the best fits to 
respective processes with no constraints of $B$ and $s_0$.

The values of parameters are given in the lower half of Table \ref{tab9}.
The $B$ and $\sqrt{s_0}$ are given by $B=0.304\pm 0.010$mb and
$\sqrt{s_0}=5.75\pm 0.34$GeV. These values are consistent with 
the results given in PDG\cite{r6}, 
$B=0.308(10)$mb and $\sqrt{s_0}$=5.38(50)GeV,
which are obtained by assuming the universality of $B$ and $s_0$.
Another interesting feature is the ratio of $Z^{ap}$.
It satisfies approximately
the quark model prediction, $Z^{\pi p}:Z^{Kp}:Z^{pp}=2:2:3$.

\section{Concluding Remarks}

\begin{figure}
\resizebox{0.5\textwidth}{!}{

  \includegraphics{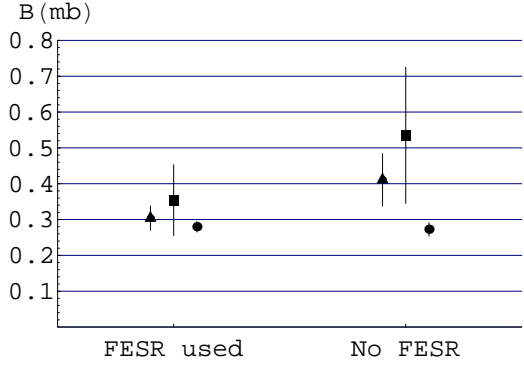}

}

\caption{Values of $B$ parameters in Table \ref{tab8}: Results with No use of FESR(RHS) and
ones with using FESR (LHS). 
Triangles, squares and circles represent $B_{\pi p}$, $B_{Kp}$ and $B_{pp}$, respectively.
Errors represent one standard($1\sigma$) deviations. 
 $B_{\pi p}=B_{Kp}=B_{pp}$ is satsified
within $1\sigma$ in the case using FESR, while we do not obtain any definite conclusion 
in the case with no use of FESR.
}

\label{fig12}

\end{figure}

In order to test the universal rise of total cross section $\sigma_{\rm tot}^{(+)}$
by log$^2s/s_0$ in all the hadron-hadron scatterings, we analyze 
$\pi^\mp p$, $K^\mp p$ and $\bar p(p)p$ scatterings independently.
Rich information of low-energy scattering data constrain, through FESR, 
the parameters in high-energy asymptotic formula to fit experimental 
$\sigma_{\rm tot}$ and $\rho$ ratios.
The values of $B$ parameters, the coefficients of log$^2s/s_0$ term in 
$\sigma_{\rm tot}^{(+)}$,
are obtained individually for three processes by these analyses.
The results are given in the LHS of Table \ref{tab8}, which is explicitly shown in Fig.~\ref{fig12}. 
We obtain
\begin{eqnarray}
&& B_{\pi p} \simeq B_{Kp} \simeq B_{pp}\ \ .\ \ \ 
\label{eq28}
\end{eqnarray}
The results are consistent with the universality of $B$ within one standard deviation. 
The universality of $B$ is suggested in our analyses.

There are several remarks in our results.

\begin{description}
\item[i)] In order to obtain the above conclusion
it is essential to use the FESR as constraints between fitting parameters.
As shown in RHS of Fig.~\ref{fig12},
if we do not use the FESR, $B_{\pi p}$ and $B_{Kp}$ cannot be estimated 
with sufficient accuracy, and we could not obatin any definite conslusion.

\item[ii)] The absolute magnitude of $B$ in the best fit is dependent fairly largely 
upon the energy region of the fittng in the present $\bar pp$ scattering data.
In the fit to all the data up to Tevatron energy $\sqrt s=1.8$TeV,
the $B_{pp}$ is estimated
as $B_{pp}=0.280\pm 0.015$mb, which
predicts the $\sigma_{\rm tot}^{pp}=108.0\pm 1.9$mb 
at the LHC energy $\sqrt s=14$TeV.
While if only the data up to the ISR energy $\sqrt s=63$GeV are 
taken into account, we obtain $B_{pp}=0.317\pm 0.034$mb, 
which predicts $\sigma_{\rm tot}^{pp}=113.2\pm 4.6$mb 
at LHC energy. 
This value is consistent with the above prediction, but its central value is 
somewhat larger than our previous result\cite{r3,r16}. 
%
%
The precise measurement of $\sigma_{\rm tot}^{pp}$ in LHC\cite{rTOTEM,rALFA} will help 
to fix the uncertainty of the absolute magnitude of $B$.

\item[iii)] Our approach can be checked by the measurement of $\sigma_{\rm tot}^{pp}$ in LHC.
Our predicted values of $\sigma_{\rm tot}^{pp}$ and $\rho^{pp}$ at LHC $\sqrt s=14$TeV
as well as the other predictions \cite{rJenk} are given in Table \ref{tablast} for comparison. 
The predictions in various models have a wide range. The LHC will select among these approaches.

\begin{table}
\begin{tabular}{l|ll}
ref. & $\sigma_{\rm tot}^{pp}$(mb) & $\rho^{pp}$\\
\hline
II[this work] & 108.0$\pm$1.9 & 0.1312$\pm$0.0024\\
II\cite{r3} & 106.3$\pm$5.1$_{\rm syst}$$\pm$2.4$_{\rm stat}$ & 0.126$\pm$0.007$_{\rm syst}$$\pm$0.004$_{\rm stat}$\\
BH\cite{r4} & 107.3$\pm$1.2 & 0.132$\pm$0.001\\
BSW\cite{rBSW} & 103.6 & 0.122\\
GLMM\cite{rGLMM,rGLM} & 92.1, 110.5 & \\
RMK\cite{rRMK} & 91.7 & \\
COMPETE\cite{rCOMPETE} & 111.5$\pm$1.2$_{\rm syst}\stackrel{+4.1}{\scriptstyle -2.1}_{\rm stat}$
  & 0.1361$\pm$0.0015$_{\rm syst}\stackrel{+0.0058}{\scriptstyle -0.0025}_{\rm stat}$\\
MN\cite{rMN} & 106.4 & 0.127\\
GKS\cite{rGKS} & 128 & 0.19\\
CS\cite{rCS} & 152 & 0.26\\
PP\cite{rPP} & 106.73$\stackrel{+7.56}{\scriptstyle -8.50}$ & 0.1378$\stackrel{+0.0042}{\scriptstyle -0.0612}$\\
ILP\cite{rILP} & 110 & 0.12\\
Landshoff\cite{rLand} & 125$\pm$25 & \\
\end{tabular}
\caption{Predictions of $\sigma_{\rm tot}^{pp}$ and $\rho^{pp}$ at LHC $\sqrt s=14$TeV in various models.}
\label{tablast}
\end{table}

\item[iv)] The best fitted values of $B_{\pi p}$ and $B_{Kp}$ are almost the same 
with each other. This result is important since it suggests 
the value of $B$ is independent of quark flavors.  
The $B_{\pi p}$ is also consistent with $B_{pp}$.
Thus, we expect the universality of 
$\sigma_{\rm meson-baryon}\simeq \sigma_{\rm baryon-baryon}$ 
independently of quark flavors in super high energies, as was suggested in 
refs.\cite{r9,r11,r12}.
In order to establish the universality of $B$ for all the hadronic scatterings
it is also important to analyze the other processes, 
such as $\Sigma^- p$ and $\Lambda p$ scatterings. 

\item[v)] If the universality of $B$ is estalished both theoretically and experimentally,
the total cross sections of all the hadronic scatterings
are described simply by Eq.~(\ref{eq3}), 
$\sigma_{\rm tot} \simeq B~{\rm log}^2s/s_0 + Z^{ap}$,
for $\sqrt s >\ \ $$\sim$$\sqrt{s_0} \simeq 5$GeV, where the effects from 
Regge poles of $P^\prime$ and of vector trajectory become negligible. 
$B$ is a universal constant. There is an interesting possibility that
$\sqrt{s_0}$ is universal.
There is no way of predicting values of $Z^{ap}$, 
which are highly non-perturbative object.  
Difference between $Z^{\pi p}$ and $Z^{Kp}$ is not large, about 3mb, and 
the $Z^{\pi p}$, $Z^{Kp}$ and $Z^{pp}$
approximately satisfy the ratio $2:2:3$, predicted by quark model.



\end{description}





%

\end{document}